\documentclass[sigconf]{acmart}
\AtBeginDocument{%
  }

\copyrightyear{2026}
\acmYear{2026}
\setcopyright{cc}
\setcctype{by}
\acmConference[RecSys '26]{20th ACM Conference on Recommender Systems}{September 27-October 02, 2026}{Minneapolis, MN, USA}
\acmBooktitle{20th ACM Conference on Recommender Systems (RecSys '26), September 27-October 02, 2026, Minneapolis, MN, USA}
\acmDOI{10.1145/3773078.3831748}
\acmISBN{979-8-4007-2284-4/2026/09}

\usepackage{enumitem}
\usepackage{float}
\usepackage{tikz}
\usetikzlibrary{shapes,arrows,positioning,calc}
\usepackage{multirow}
\usepackage{array}
\usepackage{colortbl}
\usepackage{makecell}
\usepackage{subcaption}
\usepackage{amsmath}

\newcommand{\iconlw}{0.10em}
\newcommand{\faSmile}[1][]{%
  \tikz[baseline=-0.6ex,x=1em,y=1em,line width=\iconlw,line cap=round]{%
    \draw (0,0) circle (0.42);
    \fill (-0.15,0.12) circle (0.065);
    \fill ( 0.15,0.12) circle (0.065);
    \draw (-0.20,-0.06) arc (205:335:0.22);%
  }%
}
\newcommand{\faPlayCircle}[1][]{%
  \tikz[baseline=-0.6ex,x=1em,y=1em,line width=\iconlw]{%
    \draw (0,0) circle (0.42);
    \fill (-0.11,-0.18) -- (-0.11,0.18) -- (0.21,0) -- cycle;%
  }%
}
\newcommand{\faTimes}[1][]{%
  \tikz[baseline=-0.6ex,x=1em,y=1em,line width=0.13em,line cap=round]{%
    \draw (-0.20,-0.20) -- (0.20,0.20);
    \draw (-0.20, 0.20) -- (0.20,-0.20);%
  }%
}

\definecolor{UIDens}{RGB}{24,96,168}    
\definecolor{InterMat}{RGB}{184,48,16}  
\definecolor{BidirCol}{RGB}{32,144,128} 
\definecolor{DefenseCol}{RGB}{34,139,34}   
\definecolor{DissemCol}{RGB}{100,40,160} 
\definecolor{ExtractCol}{RGB}{180,90,0}  

\colorlet{DissemFill}{DissemCol!10}
\colorlet{ExtractFill}{ExtractCol!10}
\colorlet{BidirFill}{DissemCol!8!ExtractCol!8}
\colorlet{DefenseFill}{DefenseCol!12}

\definecolor{NodeOrch}{RGB}{180,210,205}
\definecolor{NodeUser}{RGB}{200,225,222}
\definecolor{NodeItem}{RGB}{220,238,236}

\colorlet{GoodCell}{green!18!white}
\colorlet{BadCell}{black!10}
\colorlet{NACell}{black!5}

\newcommand{\dissemtag}{\colorbox{DissemFill}{\color{DissemCol}\textbf{Dissem.}}}
\newcommand{\extracttag}{\colorbox{ExtractFill}{\color{ExtractCol}\textbf{Extract.}}}

\newcommand{\defensetag}{\colorbox{DefenseFill}{\color{DefenseCol}\textbf{Defense}}}

\definecolor{HLintermat}{RGB}{255,230,230}  
\definecolor{HLtopology}{RGB}{220,240,255}   
\newcommand{\hlbox}[2]{{\setlength{\fboxsep}{1pt}\colorbox{#1}{#2}}}
\newcommand{\hlIM}[1]{\hlbox{HLintermat}{#1}}
\newcommand{\hlIT}[1]{\hlbox{HLtopology}{#1}}
\newcommand{\hlDissem}[1]{\hlbox{DissemFill}{#1}}
\newcommand{\hlExtract}[1]{\hlbox{ExtractFill}{#1}}
\newcommand{\hlDefense}[1]{\hlbox{DefenseFill}{#1}}
\newcommand{\kUI}{\hlIT{$k$}}
\newcommand{\rhoIM}{\hlIM{$\rho$}}
\newcommand{\mkUI}[1]{\colorbox{HLtopology}{$\displaystyle#1$}}
\newcommand{\mkIM}[1]{\colorbox{HLintermat}{$\displaystyle#1$}}

\makeatletter
\newcommand{\pseudosubsec}[1]{%
  \refstepcounter{subsection}%
  \subsubsection*{\thesubsection\quad #1}%
}
\makeatother

\begin{document}

\title{Attacking and Defending Multi-Agent Collaborative Filtering Systems Through Connectivity}

\author{Anjun Hu}
\email{anjun.hu@eng.ox.ac.uk}
\affiliation{%
  \institution{University of Oxford}
  \city{Oxford}
  \country{United Kingdom}
}
\author{Hanting Xie}
\email{hanting@amazon.co.uk}
\affiliation{%
  \institution{Amazon}
  \city{London}
  \country{United Kingdom}
}
\author{Saranya Govindan}
\email{saranygo@amazon.co.uk}
\affiliation{%
  \institution{Amazon}
  \city{London}
  \country{United Kingdom}
}
\author{Jas Kandola}
\email{jaskand@amazon.co.uk}
\affiliation{%
  \institution{Amazon}
  \city{London}
  \country{United Kingdom}
}
\author{Kurt Cutajar}
\email{cutajark@amazon.co.uk}
\affiliation{%
  \institution{Amazon}
  \city{London}
  \country{United Kingdom}
}

\renewcommand{\shortauthors}{Hu et al.}

\begin{abstract}
Multi-agent collaborative filtering systems coordinate autonomous LLM-powered user and item agents through natural-language interaction to refine preferences and generate recommendations. These systems inherit vulnerabilities from both their data-driven nature and multi-agent interactions, which manifest in distinct ways. Understanding how connectivity modulates vulnerability in these systems could facilitate the development of more robust recommendation pipelines.
In this work, we adapt attacks and defenses from the general multi-agent systems (MAS) literature to the agent-based CF setting, evaluating them under systematically varied connectivity in the AgentCF framework, where CF connectivity is characterized along two axes: (i) \hlIT{candidate count} (the number of item candidates per turn per user, measuring user-side interaction density) and (ii) \hlIM{catalog concentration} (the degree of item catalog overlap across users).
Our contributions include:
\textbf{(1) Adaptation:} we reproduce MAS-inspired attacks and defenses in the agentic CF domain, confirming partial transferability of original observations.
\textbf{(2) Characterization:} we characterize how the two aspects of connectivity shape attack and defense outcomes, revealing role asymmetries between user and item agents, non-monotonic temporal dynamics in attack efficacy, and divergent patterns across dissemination and extraction attack goals.
Additionally, as an exploratory extension, we assess the applicability of epidemic-inspired static metrics in ranking CF configurations by expected attack outcome, potentially enabling cost-efficient robustness assessment.
Implementation is available at  \url{https://github.com/anjunhu/ConnACF}.
\end{abstract}



\maketitle

\section{Introduction}
\label{sec:intro}

\begin{figure}[H]
\vspace{-.5em}
\centering
\resizebox{\columnwidth}{!}{%
\begin{tikzpicture}[
  unode/.style={circle, draw=none, fill=none, minimum size=0.35cm, inner sep=0, text=NodeUser!80!black},
  inode/.style={circle, draw=none, fill=none, minimum size=0.35cm, inner sep=0, text=NodeItem!60!black},
  atk/.style={circle, draw=none, fill=none, minimum size=0.35cm, inner sep=0, text=red!70!black},
  badline/.style={->, red!80!black, line width=0.4pt},
  pii/.style={font=\fontsize{4}{4}\selectfont, red!60!black, sloped, below},
  scale=1.0
]
\begin{scope}[xshift=0cm, yshift=0cm]
  \draw[rounded corners=4pt, draw=DissemCol, line width=1.2pt, fill=DissemCol!6] (-0.55,-0.95) rectangle (2.55,0.8);
  \node[font=\fontsize{4}{4}\selectfont, NodeUser!70!black] at (-0.2, 0.5) {User};
  \node[font=\fontsize{4}{4}\selectfont, NodeItem!50!black] at (1.2, 0.5) {Item};
  \node[atk]   (u1) at (-0.2, 0.25) {\faSmile[regular]}; \node[unode] (u2) at (-0.2,-0.225) {\faSmile[regular]}; \node[unode] (u3) at (-0.2,-0.7) {\faSmile[regular]};
  \node[atk]   (i1) at (1.2, 0.25) {\faPlayCircle[regular]}; \node[inode] (i2) at (1.2,-0.225) {\faPlayCircle[regular]}; \node[inode] (i3) at (1.2,-0.7) {\faPlayCircle[regular]};
  \draw[<->, DissemCol!35, line width=0.4pt] (u1)--(i1);
  \draw[<->, DissemCol!35, line width=0.4pt] (u1)--(i2);
  \draw[<->, DissemCol!35, line width=0.4pt] (u2)--(i1);
  \draw[<->, DissemCol!35, line width=0.4pt] (u2)--(i2);
  \draw[<->, DissemCol!35, line width=0.4pt] (u3)--(i2);
  \draw[<->, DissemCol!35, line width=0.4pt] (u3)--(i3);
  \draw[badline] (u1)--(i2);
  \draw[badline] (i1)--(u2) node[midway, pos=0.40, sloped, above, inner sep=0.5pt, font=\fontsize{4}{4}\selectfont, red!60!black] {\#viralmovie};
  \draw[badline] (u1)--(i3) node[midway, pos=0.5, sloped, below, inner sep=0.5pt, font=\fontsize{4}{4}\selectfont, red!60!black] {api: \{"recursive\_call"\}};
  \node[font=\scriptsize\bfseries, DissemCol] at (1.0, 0.65) {Dissemination};
  \node[font=\fontsize{4}{4}\selectfont, text=red!70!black, align=left, text width=1.0cm] at (1.95, 0.15) 
  {Spread bias or misinformation};
  \node[font=\fontsize{4}{4}\selectfont, text=red!70!black, align=left, text width=1.0cm] at (1.95, -0.5) {Exhaust system resources and interrupt service};
\end{scope}
\begin{scope}[xshift=3.3cm, yshift=0cm]
  \draw[rounded corners=4pt, draw=ExtractCol, line width=1.2pt, fill=ExtractCol!6] (-0.55,-0.95) rectangle (2.55,0.8);
  \node[unode] (u1) at (-0.2, 0.25) {\faSmile[regular]}; \node[unode] (u2) at (-0.2,-0.225) {\faSmile[regular]}; \node[unode] (u3) at (-0.2,-0.7) {\faSmile[regular]};
  \node[atk]   (i1) at (1.2, 0.25) {\faPlayCircle[regular]}; \node[inode] (i2) at (1.2,-0.225) {\faPlayCircle[regular]}; \node[inode] (i3) at (1.2,-0.7) {\faPlayCircle[regular]};
  \draw[<->, ExtractCol!35, line width=0.4pt] (u1)--(i2);
  \draw[<->, ExtractCol!35, line width=0.4pt] (u2)--(i2);
  \draw[<->, ExtractCol!35, line width=0.4pt] (u3)--(i2);
  \draw[<->, ExtractCol!35, line width=0.4pt] (u3)--(i3);
  \draw[badline] (u1)--(i1) node[midway, pos=0.4, sloped, below, inner sep=0.5pt, font=\fontsize{3.5}{3.5}\selectfont, red!60!black] {Name: John Doe};
  \draw[badline] (u2)--(i1) node[midway, pos=0.35, sloped, below, inner sep=0.5pt, font=\fontsize{3.5}{3.5}\selectfont, red!60!black] {DoB: Feb 3 2004};
  \draw[badline] (u3)--(i1) node[midway, pos=0.45, sloped, below, inner sep=0.5pt, font=\fontsize{3.5}{3.5}\selectfont, red!60!black] {Sys. Prompt: Movie Rec};
  \node[font=\scriptsize\bfseries, ExtractCol] at (1.0, 0.65) {Extraction};
  \node[font=\fontsize{4}{4}\selectfont, text=red!70!black, align=left, text width=1.0cm] at (1.95, 0.15) {Recover private user information};
  \node[font=\fontsize{4}{4}\selectfont, text=red!70!black, align=left, text width=1.0cm] at (1.95, -0.5) {Reverse engineer system prompts and topology};
\end{scope}
\begin{scope}[xshift=0cm, yshift=-1.9cm]
  \begin{scope}
    \clip[rounded corners=4pt] (-0.55,-0.95) rectangle (2.55,0.8);
    \fill[ExtractFill] (-0.55,-0.95) rectangle (1.0,0.8);
    \fill[DissemFill]  (1.0,-0.95) rectangle (2.55,0.8);
  \end{scope}
  \draw[rounded corners=4pt, draw=DissemCol!60!ExtractCol, line width=1.2pt] (-0.55,-0.95) rectangle (2.55,0.8);
  \tikzset{
    unodeS/.style={circle, draw=none, fill=none, minimum size=0.31cm, inner sep=0, text=NodeUser!80!black, font=\fontsize{6}{6}\selectfont},
    inodeS/.style={circle, draw=none, fill=none, minimum size=0.31cm, inner sep=0, text=NodeItem!60!black, font=\fontsize{6}{6}\selectfont},
    atkS/.style={circle, draw=none, fill=none, minimum size=0.31cm, inner sep=0, text=red!70!black, font=\fontsize{6}{6}\selectfont},
  }
  \node[atkS]   (lu1) at (-0.225, 0.05) {\faSmile[regular]};
  \node[unodeS] (lu2) at (-0.225,-0.30) {\faSmile[regular]};
  \node[unodeS] (lu3) at (-0.225,-0.65) {\faSmile[regular]};
  \node[atkS]   (li1) at (0.675, 0.05) {\faPlayCircle[regular]};
  \node[inodeS] (li2) at (0.675,-0.30) {\faPlayCircle[regular]};
  \node[inodeS] (li3) at (0.675,-0.65) {\faPlayCircle[regular]};
  \draw[badline] (li2)--(lu1); \draw[badline] (li3)--(lu1);
  \draw[badline] (lu2)--(li1); \draw[badline] (lu3)--(li1);
  \draw[<->, black!20, line width=0.4pt] (lu2)--(li2);
  \draw[<->, black!20, line width=0.4pt] (lu3)--(li3);
  \node[font=\fontsize{4}{4}\selectfont, text=red!70!black, align=center] at (0.225, 0.45) {$t{=}0$};
  \node[font=\fontsize{4}{4}\selectfont, text=red!70!black, align=center] at (0.225, 0.30) {reverse engineering};
  \node[atkS]   (ru1) at (1.325, 0.05) {\faSmile[regular]};
  \node[unodeS] (ru2) at (1.325,-0.30) {\faSmile[regular]};
  \node[unodeS] (ru3) at (1.325,-0.65) {\faSmile[regular]};
  \node[atkS]   (ri1) at (2.225, 0.05) {\faPlayCircle[regular]};
  \node[inodeS] (ri2) at (2.225,-0.30) {\faPlayCircle[regular]};
  \node[inodeS] (ri3) at (2.225,-0.65) {\faPlayCircle[regular]};
  \draw[badline] (ru1)--(ri2); \draw[badline] (ru1)--(ri3);
  \draw[badline] (ri1)--(ru2); \draw[badline] (ri1)--(ru3);
  \draw[<->, black!20, line width=0.4pt] (ru2)--(ri2);
  \draw[<->, black!20, line width=0.4pt] (ru3)--(ri3);
  \node[font=\fontsize{4}{4}\selectfont, text=red!70!black, align=center] at (1.775, 0.45) {$t{=}2$};
  \node[font=\fontsize{4}{4}\selectfont, text=red!70!black, align=center] at (1.775, 0.30) { bias dissemination};
  \node[font=\scriptsize\bfseries] at (1.0, 0.65) {\textcolor{ExtractCol}{Bidirec}\textcolor{DissemCol}{tional}};
\end{scope}
\begin{scope}[xshift=3.3cm, yshift=-1.9cm]
  \draw[rounded corners=4pt, draw=DefenseCol, line width=1.2pt, fill=DefenseFill] (-0.55,-0.95) rectangle (2.55,0.8);
  \tikzset{
      unodeS/.style={circle, draw=none, fill=none, minimum size=0.31cm, inner sep=0, text=NodeUser!80!black, font=\fontsize{6}{6}\selectfont},
      inodeS/.style={circle, draw=none, fill=none, minimum size=0.31cm, inner sep=0, text=NodeItem!60!black, font=\fontsize{6}{6}\selectfont},
      atkS/.style={circle, draw=none, fill=none, minimum size=0.31cm, inner sep=0, text=red!70!black, font=\fontsize{6}{6}\selectfont},
    }
  \node[atkS]   (lu1) at (-0.225, 0.05) {\faSmile[regular]};
  \node[unodeS] (lu2) at (-0.225,-0.30) {\faSmile[regular]};
  \node[unodeS] (lu3) at (-0.225,-0.65) {\faSmile[regular]};
  \node[atkS]   (li1) at (0.675, 0.05) {\faPlayCircle[regular]};
  \node[inodeS] (li2) at (0.675,-0.30) {\faPlayCircle[regular]};
  \node[inodeS] (li3) at (0.675,-0.65) {\faPlayCircle[regular]};
  \draw[<->, DefenseCol!35, line width=0.4pt] (lu2)--(li2);
  \draw[<->, DefenseCol!35, line width=0.4pt] (lu3)--(li3);
  \draw[badline] (lu1)--(li2); \draw[badline] (lu1)--(li3);
  \draw[badline] (li1)--(lu2); \draw[badline] (li1)--(lu3);
  \node[font=\fontsize{4}{4}\selectfont, DefenseCol!80!black, align=center] at (0.225, 0.45) {$t{=}0$};
  \node[font=\fontsize{4}{4}\selectfont, DefenseCol!80!black, align=center] at (0.225, 0.30) {audit and train};

  \node[atkS]   (ru1) at (1.325, 0.05) {\faSmile[regular]};
  \node[unodeS] (ru2) at (1.325,-0.30) {\faSmile[regular]};
  \node[unodeS] (ru3) at (1.325,-0.65) {\faSmile[regular]};
  \node[atkS]   (ri1) at (2.225, 0.05) {\faPlayCircle[regular]};
  \node[inodeS] (ri2) at (2.225,-0.30) {\faPlayCircle[regular]};
  \node[inodeS] (ri3) at (2.225,-0.65) {\faPlayCircle[regular]};
  \draw[<->, DefenseCol!35, line width=0.4pt] (ru2)--(ri2);
  \draw[<->, DefenseCol!35, line width=0.4pt] (ru3)--(ri3);
  \draw[->, red!60!black!50, dashed, line width=0.4pt] (ri1)--(ru2);
  \draw[->, red!60!black!50, dashed, line width=0.4pt] (ri1)--(ru3);
  \draw[->, red!60!black!50, dashed, line width=0.4pt] (ru1)--(ri2);
  \draw[->, red!60!black!50, dashed, line width=0.4pt] (ru1)--(ri3);
  \node[font=\fontsize{6}{6}\selectfont, DefenseCol] at (1.325, 0.05) {\faTimes};
  \node[font=\fontsize{6}{6}\selectfont, DefenseCol] at (2.225, 0.05) {\faTimes};
  \node[font=\fontsize{4}{4}\selectfont, DefenseCol!80!black, align=center] at (1.775, 0.45) {$t{=}30$};
  \node[font=\fontsize{4}{4}\selectfont, DefenseCol!80!black, align=center] at (1.775, 0.30) {defend and prune};
  \node[font=\scriptsize\bfseries, DefenseCol] at (1.0, 0.65) {Defense};
\end{scope}
\end{tikzpicture}
}
\caption{
Attack/defense families reproduced on multi-agent CF. Red nodes are attackers; layouts are schematic. How connectivity modulates the outcome is the subject of this work.
\vspace{-1em}
}
\label{fig:overview}
\end{figure}
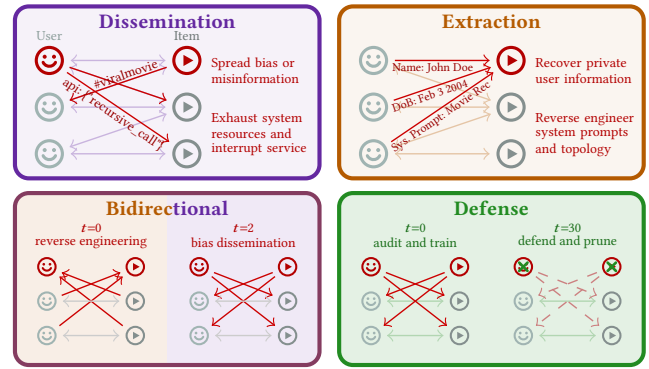

Multi-agent collaborative filtering (CF) systems augment traditional recommendation with stateful, autonomous LLM-powered agents that represent users and items~\cite{liu2024agentcf}.
Connectivity in these systems arises from \hlIT{structural design choices} (communication topology and candidate item count per round) and \hlIM{users' historical interactions} (overlap in past interactions and preference alignment).
These dimensions shape both the evolution of preferences and how vulnerabilities propagate through the system. The latter has received limited attention, and this work aims to characterize the connectivity-robustness relationship in multi-agent CF.

The same inter-agent communication pathways that enable preference refinement could also introduce (non-exhaustively) two types of unwanted information flows:
Firstly, unwanted \emph{dissemination} may occur when a compromised agent injects a payload that spreads outward: a poisoned item description persists in a user's memory, shapes future interactions, and reaches peers through opinion exchange.
Secondly, unwanted \emph{extraction} may occur when an attacker agent recovers unauthorized proprietary information stored in a victim agent's memory, with leakage observable on communication channels accumulating across interaction turns.
Both directions are potentially governed by the same connectivity variables, resulting in variations in recommendation quality and system safety.

Recommender systems are known to be susceptible to data poisoning and adversarial attacks~\cite{deldjoo2021adversarial_survey,nguyen2024manipulating}. Recent red-teaming works increasingly target agentic recommenders through their stateful nature and inter-agent communication channels, devising novel attacks through profile poisoning and memory corruption~\cite{ning2024CheatAgent,yang2025drunkagent,zhang2024stealthy}.
These attacks confirm system vulnerability but are tested on fixed system configurations, leaving open whether attack efficacy changes as connectivity varies.
On the other hand, general-domain MAS research has shown connectivity to be a significant determinant of attack outcomes~\cite{liang2025dont,liu2025topology,zhou2025corba,yu2024netsafe}, but those findings come from systems that differ substantially in scale and structure from CF, and their transferability is not yet established.

Multi-agent CF has structural properties that are distinct from query-driven MAS, making attack transferability non-trivial.
(1) CF systems are \emph{bipartite}, with asymmetric user and item roles that have no direct counterpart in role-homogeneous or tool-heterogeneous MAS networks.
(2) Connectivity manifests more diversely in CF, originating from both inference system design and historical interaction data, defined more formally in \S\ref{subsubsec:two-connectivity}. 
(3) CF usually operates at larger scale with agent states persisting over longer interaction horizons, giving rise to intriguing temporal dynamics (F3, \S\ref{sec:evaluation}) not previously reported for smaller task-driven MAS.

In this work, we reproduce topology-vulnerability findings from general MAS in the CF setting, using AgentCF~\cite{liu2024agentcf} as the primary evaluation platform and incorporating MACF~\cite{xia2025multiagent}.
We scope our threat model to application-layer adversaries who compromise a fraction of user and/or item agents through prompt injection or profile poisoning, with goals spanning dissemination (preference manipulation or resource exhaustion), extraction (privacy breach or reverse engineering), or a combination of both.
\noindent Our contributions include:
\textbf{(1) Adaptation and Reproduction.}
We adapt MAS attacks and defenses to the multi-agent CF setting and validate the applicability of original methods.
\textbf{(2) Characterizing the Effect of Connectivity.}
We find that candidate count and catalog concentration produce qualitatively different effects on attack success, with direction and magnitude varying across attacker goals and role configurations, revealing user-item partition asymmetries unique to CF systems.
Additionally, we assess the applicability of epidemic-inspired static metrics derived from system configuration parameters in ranking CF configurations by expected attack outcome, potentially enabling cost-efficient, \textit{a priori} risk assessment.

\section{Related Work}
\label{sec:related-work}

Multi-agent recommender systems (MARS) inherit vulnerabilities from both their data-driven nature and their multi-agent communication structure.
Our work sits at the intersection of MAS and RecSys safety, bringing topology awareness from MAS works to MARS and grounding evaluation in CF's unique structures.
\vspace{2.3pt}\\
\noindent\textbf{Multi-Agent Recommender Systems.} LLM-based recommender systems have evolved from single-agent interfaces~\cite{2024recai,wang2023recmind,huang2023recommender,zhang2024toolrec} towards multi-agent architectures~\cite{maragheh2025the,huang2025tutorial,huang2025agenticrecommendersystemsera,peng2025a} with diverse topologies and configurations: \textit{centralized} systems route coordination through a hub or orchestrator~\cite{wang2024MACRec,yu2025thoughtaugmented,MAS4POI2024}; \textit{mesh} systems model recommendation as emergent peer-to-peer behavior~\cite{liu2024agentcf,liu2025agentcf,xia2025multiagent,zhang2024prospect} and facilitate behavioral simulation~\cite{wang2023recagent,zhang2024generative,zhu2024a}; and \textit{hybrid} systems combine both~\cite{shi2024billp,thakkar2024personalized}. Their safety is receiving increasing attention~\cite{hui2025toward}, and our work builds upon this line of research.
\vspace{2.3pt}\\
\noindent\textbf{LLM and Agentic Recommender System Safety.}
Classical poisoning and adversarial attacks spanning item-profile manipulation, shilling, and prompt injection~\cite{lin2022shilling,IndirectAD,wang2024poisoning,wu2023attackingpretrainedrecommendation,zhang2024stealthy,deldjoo2021adversarial_survey,nguyen2024manipulating} remain relevant to LLM and agentic recommenders. Recent work targets agentic recommenders directly via profile poisoning, memory corruption, backdoor implantation, and retrieval poisoning~\cite{ning2024CheatAgent,yang2025drunkagent,ning2025exploring,wang2025shilling,li2025semanticshield,zhang2024humanimperceptible,yang2025retrievalaugmented,wang2024idfree}. Indirect prompt injection and self-replicating payloads further extend attack surfaces to inter-agent communication~\cite{zhan2024injecagent,lee2024prompt}. Control-flow hijacking demonstrates the blast radius of these vulnerabilities in general MAS~\cite{jha2025breaking,triedman2025multiagent}. Privacy leakage~\cite{wang2025privacy_inversion,patil2025the,yagoubi2026agentleak,zhang2021membership}, collusion~\cite{ghaemi2025collusion,huang2026emergent}, collective manipulation~\cite{toni2026with}, and closed-loop ranking exploitation~\cite{cheng2026let} represent further threat dimensions. We adapt representative attacks from this body of work to the CF setting, and ask how their efficacy changes as connectivity varies.
\vspace{2.3pt}\\
\noindent\textbf{Evaluating MAS Safety.}
ASB~\cite{asb2025} and TAMAS~\cite{kavathekar2025tamas} benchmark attacks and defenses in general MAS; AgentLeak~\cite{yagoubi2026agentleak} and failure-mode analyses~\cite{cemri2025why} cover privacy leakage and systematic breakdowns. In RecSys, safety tutorials~\cite{recsys23_trustworthy,wsdm25_secure}, AgentRecBench~\cite{shang2025agentrecbench}, and agentic evaluation at scale~\cite{zhang2025nohuman} survey the broader landscape. Evaluation has also expanded beyond utility~\cite{jiang2024beyond} to consider safety and trustworthiness. Our evaluation harness is informed by these benchmarks and extends them with per-partition, per-turn metrics that separate transient from steady-state regimes.
\vspace{2.3pt}\\
\noindent\textbf{Topology Awareness in MAS Safety.}
NetSafe~\cite{yu2024netsafe} demonstrates that lower-connectivity topologies are safer and introduces static graph metrics predicting safety rankings. Stealthy multi-round tampering~\cite{yan2025attack}, optimized prompt attacks~\cite{khan2025textitagents}, and information propagation analysis~\cite{shen2025understanding} further characterize how topology shapes attack spread. 
GNN-based defenses~\cite{wang2025gsafeguard,pan2025explainable,miao2025blindguard,zhou2025guardian,feng2025sentinelnet} operationalize topology awareness on the defense side, several of which we adapt and evaluate in this work. 
We bring topology awareness to CF by reproducing a dual-source connectivity-driven evaluation, characterizing how candidate count and catalog concentration jointly shape outcomes across user and item partitions.

\section{Experimental Setup}
\label{sec:setup}

We build a unified evaluation harness on top of AgentCF~\cite{liu2024agentcf}, also incorporating MACF~\cite{xia2025multiagent}, equipping both with a common attack, defence and metric tracking surface and exposing two connectivity axes as independent experimental variables.

\pseudosubsec{Base Systems.}\label{subsubsec:base-sys-agentcf-macf}
AgentCF~\cite{liu2024agentcf} is a bipartite mesh CF system: despite being orchestrator-moderated, user and item agents can directly impact each others' memory state. In the canonical setting, at each step an orchestrator selects a (user, positive item, negative item) triple; the user ranks the two items (\hlIT{$k$=2}); both item agents update their descriptions; and the user updates its preference memory. An optional user-user opinion-exchange step lets users share preference summaries with peers. This canonical setting corresponds to \hlIT{$k$=2} and a \hlIM{medium-overlap catalog}. We generalize this system to other \kUI{} by changing the binary task to 3-item ranking in case of \hlIT{$k$=3} and binary like/dislike feedback in case of \hlIT{$k$=1}; and vary \hlIM{catalog density} by resampling the dataset (\S\ref{subsubsec:dataset-task}).
MACF~\cite{xia2025multiagent} is an orchestrator-mediated star topology CF system (\hlIT{$k{=}0$}): all user-item communication is routed through a central hub and no direct U-I exchange triggers immediate memory updates. MACF is included in the linked codebase and available for analysis, but excluded from the plots in \S\ref{sec:evaluation} since its workflow introduces structural confounds beyond connectivity. Moreover, attacks with extraction components are not applicable to MACF without significant changes due to the lack of immediate peripheral agent contact.

\pseudosubsec{Memory Management.}\label{subsubsec:memory}
Each agent's profile is a rolling episodic memory list: after each interaction the LLM appends a semantic summary, providing a persistent structured surface for injection (CORBA, NetSafe, etc.) and extraction (MAMA, MASLeak, etc.).

\pseudosubsec{Two Connectivity Axes}\label{subsubsec:two-connectivity}
The two axes are illustrated in Fig.~\ref{fig:agentcf_variants}. \kUI{} is a system configuration set at inference time independently of the dataset; \rhoIM{} is determined by historical interaction data.\\
{(1) \hlIT{Candidate count} (\kUI{}, inference-time communication density).}
$k\in \{1, 2, 3\}$ is the number of items each user can interact with at each turn. In other words, it is the user's decision bandwidth. A higher \kUI{} means each user can immediately impact more items.

\noindent {(2) \hlIM{Catalog concentration} (\rhoIM{}, interaction history density).}
We fix $n_U$=100 users and vary item catalog size $n_I$ by re-sampling MovieLens-100K (\S\ref{subsubsec:dataset-task}) to $n_I \in \{200, 100, 50\}$, giving $\rho \equiv n_U/n_I \in \{0.5, 1, 2\}$. Denser catalogs concentrate the same number of historical interactions on fewer items, making each item more influential.

\definecolor{Wire1c}{RGB}{183,47,20}
\definecolor{Wire2c}{RGB}{32,151,131}
\definecolor{Wire3c}{RGB}{24,96,169}

\begin{figure}[h]
\centering
\resizebox{\columnwidth}{!}{%
\begin{tikzpicture}[
    agent/.style={rectangle, draw=black!30, rounded corners, minimum size=0.7cm, font=\small\bfseries, text=black},
    orch/.style={agent, fill=NodeOrch},
    user/.style={agent, fill=NodeUser},
    item/.style={agent, fill=NodeItem},
    ctrl/.style={-, black!40, thin},
    w1/.style={-, Wire1c, line width=1.8pt},
    w2/.style={-, Wire2c, line width=1.8pt},
    w3/.style={-, Wire3c, line width=1.8pt},
    lbl/.style={font=\scriptsize, align=center},
    scale=0.85, transform shape
]
\begin{scope}[xshift=0cm]
  \node[orch] (H0) at (0,0) {$A_0$};
  \foreach \u/\a in {U1/150,U2/90,U3/30}  \node[user] (\u) at (\a:1.1) {\u};
  \foreach \i/\a in {I1/210,I2/270,I3/330} \node[item] (\i) at (\a:1.1) {\i};
  \foreach \x in {U1,U2,U3,I1,I2,I3} \draw[ctrl] (H0)--(\x);
  \node[lbl] at (0,-1.9) {$k{=}0$\\(MACF)};
\end{scope}
\begin{scope}[xshift=3.2cm]
  \node[orch] (H1) at (0,0) {$A_0$};
  \foreach \u/\a in {U1/150,U2/90,U3/30}  \node[user] (\u) at (\a:1.1) {\u};
  \foreach \i/\a in {I1/210,I2/270,I3/330} \node[item] (\i) at (\a:1.1) {\i};
  \foreach \x in {U1,U2,U3,I1,I2,I3} \draw[ctrl] (H1)--(\x);
  \draw[w1] (U1)--(I1); \draw[w1] (U2)--(I2); \draw[w1] (U3)--(I3);
  \node[lbl] at (0,-1.9) {$k{=}1$};
\end{scope}
\begin{scope}[xshift=6.4cm]
  \node[orch] (H2) at (0,0) {$A_0$};
  \foreach \u/\a in {U1/150,U2/90,U3/30}  \node[user] (\u) at (\a:1.1) {\u};
  \foreach \i/\a in {I1/210,I2/270,I3/330} \node[item] (\i) at (\a:1.1) {\i};
  \foreach \x in {U1,U2,U3,I1,I2,I3} \draw[ctrl] (H2)--(\x);
  \draw[w2] (U1)--(I1); \draw[w2] (U1)--(I2);
  \draw[w2] (U2)--(I2); \draw[w2] (U2)--(I3);
  \draw[w2] (U3)--(I1); \draw[w2] (U3)--(I3);
  \node[lbl] at (0,-1.9) {$k{=}2$\\(AgentCF)};
\end{scope}
\begin{scope}[xshift=9.6cm]
  \node[orch] (H3) at (0,0) {$A_0$};
  \foreach \u/\a in {U1/150,U2/90,U3/30}  \node[user] (\u) at (\a:1.1) {\u};
  \foreach \i/\a in {I1/210,I2/270,I3/330} \node[item] (\i) at (\a:1.1) {\i};
  \foreach \x in {U1,U2,U3,I1,I2,I3} \draw[ctrl] (H3)--(\x);
  \foreach \u in {U1,U2,U3} \foreach \i in {I1,I2,I3} \draw[w3] (\u)--(\i);
  \node[lbl] at (0,-1.9) {$k{=}3$};
\end{scope}
\begin{scope}[yshift=-2.5cm, xshift=-0.7cm]
  \node[orch, minimum size=0.35cm] at (0,0) {};
  \node[font=\small, right=0.1cm] at (0.18,0) {Orch.};
  \node[user, minimum size=0.35cm] at (2.0,0) {};
  \node[font=\small, right=0.1cm] at (2.18,0) {User};
  \node[item, minimum size=0.35cm, draw=black!30] at (4.0,0) {};
  \node[font=\small, right=0.1cm] at (4.18,0) {Item};
  \draw[w1,thick] (5.8,-0.0)--(6.3,-0.0); \node[font=\small] at (6.95,0) {$k{=}1$};
  \draw[w2,thick] (7.8,-0.0)--(8.3,-0.0); \node[font=\small] at (8.95,0) {$k{=}2$};
  \draw[w3,thick] (9.8,-0.0)--(10.3,-0.0); \node[font=\small] at (10.95,0) {$k{=}3$};
\end{scope}
\end{tikzpicture}
}
\includegraphics[width=\columnwidth]{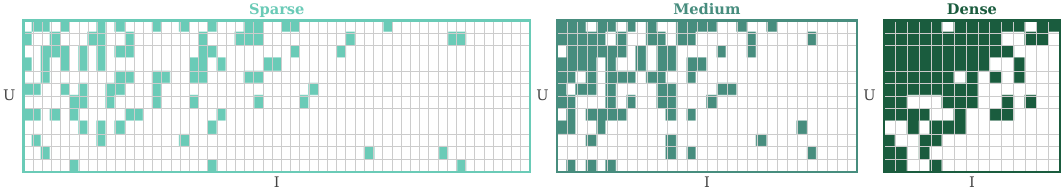}
\caption{(Top) \hlIT{Candidate count} variants: graph edge colors match \kUI{}=\textcolor[RGB]{183,47,20}{1}/\textcolor[RGB]{32,151,131}{2}/\textcolor[RGB]{24,96,169}{3} line colors used in metric plots. (Bottom) \hlIM{catalog concentration} variants: \rhoIM{}=\textcolor[RGB]{106,203,183}{\textbf{0.5}}/\textcolor[RGB]{71,141,126}{\textbf{1}}/\textcolor[RGB]{26,92,62}{\textbf{2}} correspond to low/medium/high overlap across users.}
\label{fig:agentcf_variants}
\end{figure}
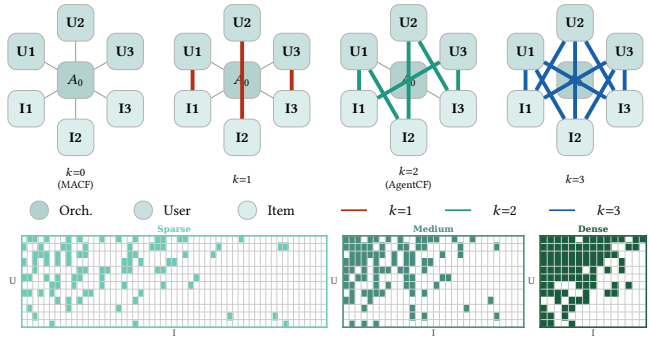

\pseudosubsec{Dataset and Task.}\label{subsubsec:dataset-task}
We use the MovieLens-100K dataset~\cite{harper2015movielens}, subsampling 100 users and varying the item catalog size to produce three \hlIM{catalog concentration} levels (\S\ref{subsubsec:two-connectivity}). MovieLens-100K is the canonical benchmark of the AgentCF base system we reproduce~\cite{liu2024agentcf}; keeping to it lets us faithfully replicate the original system and isolate connectivity effects rather than confounding them with a change of domain. The single-dataset, single-domain scope is a deliberate reproduction choice and a stated limitation (\S\ref{sec:conclusion}).
The task is sequential movie rating prediction: at each turn, each user agent takes an action (rate/choose/rank) on $k\in \{1, 2, 3\}$ items, updating its preference memory and the item descriptions accordingly.
In parallel to profile optimization, the system task (sequential item ranking prediction) is triggered by separate evaluation queries.

\pseudosubsec{Threat Models.}\label{subsubsec:threat-models}
We consider application-layer adversaries who compromise a fixed fraction $\alpha$ of user agents, item agents, or both ($\alpha_U$, $\alpha_I$), with $(0.25, 0.25)$ for main experiments and more combinations as ablations (\S\ref{subsubsec:ablation-concentration}-\ref{subsubsec:ablation-role-split}).
Compromise is achieved via two surfaces: (i) \emph{system-prompt override} (replacing an agent's system/task prompt with an adversarial persona) and (ii) \emph{user profile/item description poisoning} (seeding canary content into a user agent's memory or an item agent's description).
LLM weights and inference infrastructure are assumed secure; the adversary has no access to model internals or the orchestrator's scheduling logic.
Attacker goals fall into two directions:
\begin{itemize}[nosep,leftmargin=*]
\item \textbf{Dissemination} (bias induction, resource exhaustion). Attacker agents inject canary concepts (e.g., esoteric theatrical genres such as Letterist cinema) or self-replicating blocking payloads into their profiles. Success is the \emph{contamination rate}, the fraction of non-attacker agents whose current memory contains these payloads, tracked per turn for the user ($U$) and item ($I$) partitions.

\item \textbf{Extraction} (privacy breach, reverse engineering). Attacker agents elicit private information from victim memories through targeted dialogues. Success is the \emph{leakage rate}: for example, the fraction of seeded PII fields recovered; or the fraction of system-design IP (prompts, instructions, topology) recovered.
\end{itemize}

\pseudosubsec{Metrics: Attack Success Rate (ASR) via LLM Judge.}\label{subsubsec:metrics-asr-llmjudge}
For each turn, an LLM judge evaluates every non-attacker agent by comparing its current memory against its original state and the known canary concepts, returning a binary contaminated/clean label. 
ASR is defined as the ratio between contaminated non-attacker agents and total non-attacker agents, reported separately for user ($U$) and item ($I$) partitions per turn.
ASR percentages exclude attacker agents, so reported rates reflect spread or leakage among the benign population only.
For canary-style dissemination and PII-extraction attacks, where the seeded concept or field is known a priori, contamination and leakage can also be checked deterministically via exact and embedding-based matching; we use these deterministic checks to spot-check the judge's labels and treat the LLM judge as the primary instrument. We flag full human-annotated calibration of the judge (false-positive/false-negative rates, prompt-sensitivity analysis) as a limitation and a target for extension.
For temporal regime-aware analysis, we also distinguish the initial transient stage \textbf{slope} ($U_{\text{tr}}$, $I_{\text{tr}}$) from the steady-state \textbf{mean} ($U_{\text{ss}}$, $I_{\text{ss}}$) in \S\ref{sec:evaluation}. Attack-specific metrics (e.g., blocking or Denial-of-Service rate, per-category PII leakage, topology recovery rate) are reported alongside ASR where applicable.

\begin{figure*}[t]
\centering
\vspace{-0.5em}
\resizebox{\columnwidth}{!}{%
\begin{tikzpicture}
  \draw[line width=1.8pt,color={rgb,255:red,183;green,47;blue,20}]  (0,0)--(0.6,0);   \node[color={rgb,255:red,183;green,47;blue,20}] at (0.3,0) {\scalebox{1.15}{\ding{117}}};  \node[font=\small,right=0.05cm] at (0.6,0)  {$k{=}1$, $\rho{=}1$};
  \draw[line width=1.8pt,color={rgb,255:red,32;green,151;blue,131}] (2.5,0)--(3.1,0); \node[color={rgb,255:red,32;green,151;blue,131}] at (2.8,0) {\scalebox{1.5}{$\boldsymbol{\times}$}};    \node[font=\small,right=0.05cm] at (3.1,0)  {$k{=}2$, $\rho{=}1$};
  \draw[line width=1.8pt,color={rgb,255:red,24;green,96;blue,169}]  (5.0,0)--(5.6,0); \node[color={rgb,255:red,24;green,96;blue,169}]  at (5.3,0) {\tikz[baseline=-0.5ex]{\fill[color={rgb,255:red,24;green,96;blue,169}] (-0.11,0.13)--(0.11,0.13)--(0,0)--cycle (-0.11,-0.13)--(0.11,-0.13)--(0,0)--cycle;}};  \node[font=\small,right=0.05cm] at (5.6,0)  {$k{=}3$, $\rho{=}1$};
  \draw[line width=1.8pt,color={rgb,255:red,106;green,203;blue,183}](10.2,0)--(10.8,0); \node[color={rgb,255:red,106;green,203;blue,183}] at (10.5,0) {\scalebox{1.15}{\ding{117}}};  \node[font=\small,right=0.05cm] at (10.8,0)  {$k{=}2$, $\rho{=}0.5$};
  \draw[line width=1.8pt,color={rgb,255:red,71;green,141;blue,126}] (12.7,0)--(13.3,0); \node[color={rgb,255:red,71;green,141;blue,126}]  at (13.0,0) {\scalebox{1.5}{$\boldsymbol{\times}$}};    \node[font=\small,right=0.05cm] at (13.3,0)  {$k{=}2$, $\rho{=}1$};
  \draw[line width=1.8pt,color={rgb,255:red,26;green,92;blue,62}]   (15.2,0)--(15.8,0);\node[color={rgb,255:red,26;green,92;blue,62}]  at (15.5,0){\tikz[baseline=-0.5ex]{\fill[color={rgb,255:red,26;green,92;blue,62}] (-0.11,0.13)--(0.11,0.13)--(0,0)--cycle (-0.11,-0.13)--(0.11,-0.13)--(0,0)--cycle;}};  \node[font=\small,right=0.05cm] at (15.8,0) {$k{=}2$, $\rho{=}2$};
\end{tikzpicture}%
}
\vspace{-1.4em}
\end{figure*}

\begin{figure*}[t]
  \centering
  \begin{minipage}[t]{0.48\linewidth}
      \centering
      \includegraphics[width=\columnwidth]{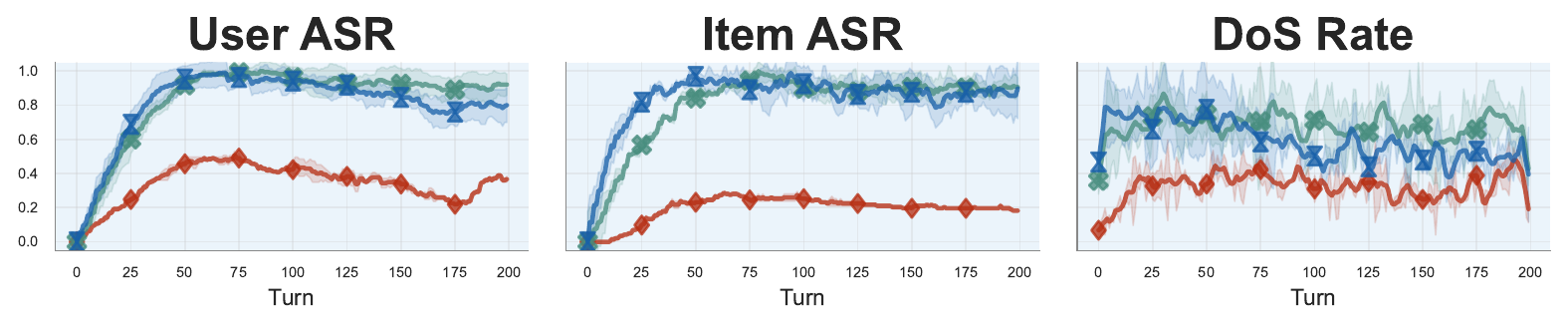}
      \includegraphics[width=\columnwidth]{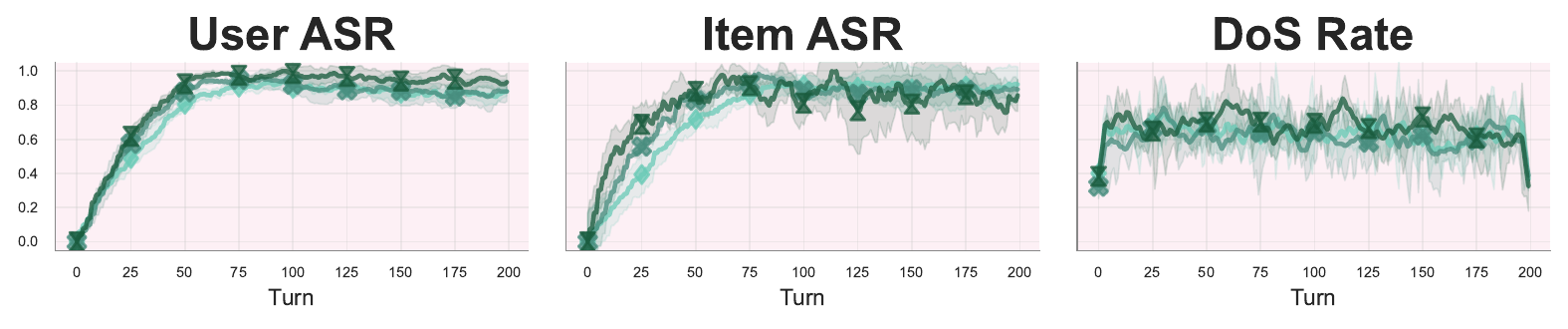}
      \vspace{-2em}
      \caption{CORBA ASR and DoS rate (y-axis) across turns (x-axis) and connectivity configurations:  \hlIT{candidate count} ($k$, top) and \hlIM{catalog concentration} ($\rho$, bottom).}
      \label{fig:corba_modulation}
  \end{minipage}%
  \hfill
  \begin{minipage}[t]{0.48\linewidth}
      \centering
      \includegraphics[width=\columnwidth, clip, trim=0 0 0 0pt]{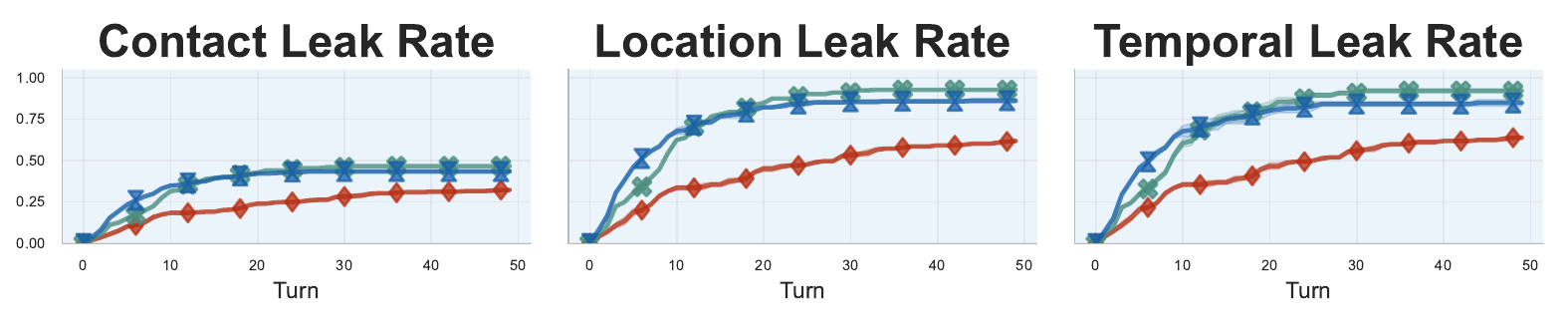}
      \includegraphics[width=\columnwidth, clip, trim=0 0 0 0pt]{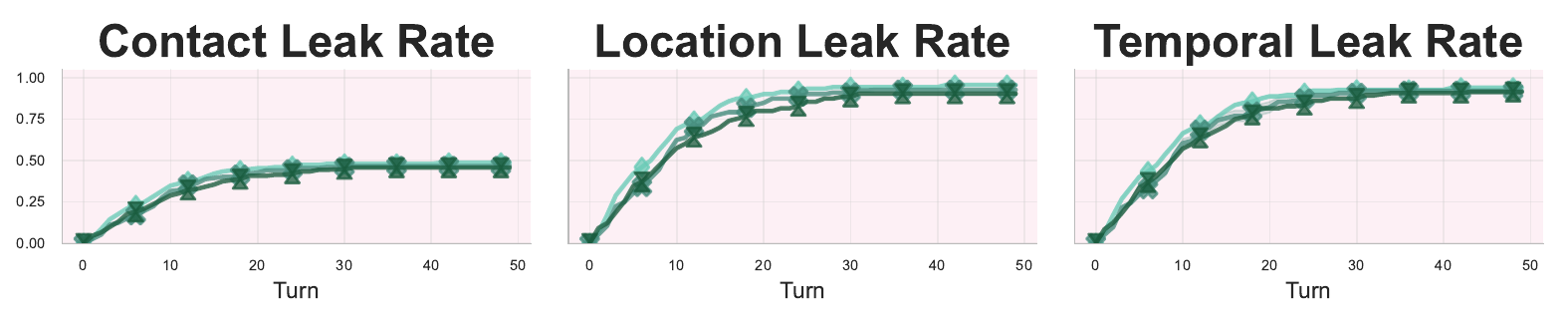}
      \vspace{-2em}
      \caption{MAMA per-category PII leakage across various \hlIT{$k$} and \hlIM{$\rho$} connectivity configurations.}
    \label{fig:mama_modulation}
  \end{minipage}
\end{figure*}

\pseudosubsec{Computational Environment.}\label{subsubsec:compute-env}
All experiments were conducted on an AWS EC2 {g6.12xlarge} instance using the Deep Learning OSS Nvidia Driver AMI image.
The instance is equipped with 4 NVIDIA L4 GPUs (23\,GB VRAM each), running Ubuntu 24.04.4 LTS (kernel 6.17.0-1010-aws), CUDA 12.9 and PyTorch 2.11. Default agent LLM is Qwen3-235B-A22B. Default judge model is Claude Sonnet 4.5.
Full conda environment specification and run scripts are included in the repository.
LLM inference is handled via Amazon Bedrock. 

\section{Evaluation}
\label{sec:evaluation}

This section presents how we adapt existing MAS evaluation, attack and defenses to the CF setting and reproduced results on connectivity variants.
We reproduced red-teaming methods grouped by the direction of unwanted information flow they cause: \text{dissemination}-driven (\S\ref{subsec:findings-dissemination}), \text{extraction}-driven (\S\ref{subsec:findings-extraction}), and \text{bidirectional} (\S\ref{subsec:findings-bidirectional}). Defenses follow in \S\ref{subsec:findings-defenses}. 
For each method, we report: \textbf{Adaptation} (original methods, metrics, how we adapt them to CF setting) and \textbf{Characterization} (per-turn temporal metrics that demonstrate how the two connectivity axes affect efficacy). 

\vspace{2pt}
\noindent
Three cross-cutting patterns are observed across adapted methods:

\noindent\textbf{(F1) ASRs exhibit growth-saturation patterns.}
Contamination and leakage curves exhibit rapid early growth that decelerates and saturates, consistent with epidemic spreading dynamics observed in prior MAS red-teaming work.

\noindent\textbf{(F2) Role-based asymmetry.}
For attacks that are applicable to both User and Item roles, attack efficacy on the two agentic role partitions ($U_{**}$ and $I_{**}$) and the modulation effects of both connectivity axes are generally not the same and warrant separate tracking.

\noindent\textbf{(F3) Decoupled slope and plateau.}
A faster early-phase growth rate (higher initial transient slope, $U_{\text{tr}}$, $I_{\text{tr}}$) does not always translate to higher steady-state mean attack performance ($U_{\text{ss}}$, $I_{\text{ss}}$).

\subsection{Dissemination-Driven Attacks}
\label{subsec:findings-dissemination}
We first reproduce two dissemination-driven attacks: CORBA (resource exhaustion via a self-replicating recursive blocking payload) and NetSafe (bias and misinformation spread).
Increased connectivity in both \kUI{} and \rhoIM{} generally facilitate dissemination, with attack-specific nuances detailed below.

\subsubsection{\textcolor{DissemCol}{\hlDissem{\textbf{CORBA}}}}
\label{subsubsec:corba}

\textbf{Adaptation.}
CORBA~\cite{zhou2025corba} injects a self-replicating recursive blocking payload that combines contagion (pass to neighbours) with recursion (re-inject to oneself), achieving near-complete contamination within 20 turns. Our adapted version seeds the payload into a target item agent's memory and into the system prompts of a fixed fraction of user agents. Since AgentCF's backward reflection step can partially wash out the payload, lower absolute rates than the original are expected. We track user and item victim rate separately, along with overall Denial-of-Service (DoS) rate.

\noindent \textbf{Characterization.}
As shown in Fig.~\ref{fig:corba_modulation}, both partitions follow growth-saturation patterns  (F1), but respond differently to each axis (F2). Increasing connectivity in \kUI{} raises both user and item initial growth slopes; at steady state, $k{=}2$ and $k{=}3$ curves converge while both remain ahead of $k{=}1$. Varying \rhoIM{} leaves user-side non-monotonic throughout. Meanwhile, higher \rhoIM{} yields higher item-side initial ASR growth slope, but item-side ASR becomes non-monotonic at steady state (F3).

\subsubsection{\textcolor{DissemCol}{\hlDissem{\textbf{NetSafe}}}}
\label{subsubsec:netsafe}
\textbf{Adaptation.}
NetSafe~\cite{yu2024netsafe} replaces the system prompts of a fixed fraction of agents with attackers' and measures contamination rate across the network, finding that higher connectivity increases adversarial spread and that static attacker-distance metrics predict safety rankings better than classical graph metrics. We adapt the misinformation variant by seeding attacker agents via both surfaces: a poisoned system prompt and profile for user agents, or a poisoned product description for item agents. In addition to the default setting, three ablations (attacker ratio, role-split and style variants, and LLM swap) are reported in \S\ref{sec:ablation}.

\noindent \textbf{Characterization.}
As shown in Fig.~\ref{subfig:ablation-rolesplit-netsafe}, both partitions follow growth-saturation patterns (F1) but respond differently to connectivity modulation (F2). User-side contamination is non-monotonic with \kUI{} across both transient and steady-state regimes. Item-side transient slope is also non-monotonic ($I_{\text{tr}}^{k=3} > I_{\text{tr}}^{k=1} > I_{\text{tr}}^{k=2}$), but a reversal between $k{=}1$ and $k{=}2$ eventually resolves: higher \kUI{} yields higher item-side plateaus (F3). Varying \rhoIM{} leaves user-side non-monotonic throughout, while higher \rhoIM{} yields higher item-side growth rates and final contamination levels.

\begin{figure*}[t]
\centering
\vspace{-0.5em}
\resizebox{\columnwidth}{!}{%
\begin{tikzpicture}
  \draw[line width=1.8pt,color={rgb,255:red,183;green,47;blue,20}]  (0,0)--(0.6,0);   \node[color={rgb,255:red,183;green,47;blue,20}] at (0.3,0) {\scalebox{1.15}{\ding{117}}};  \node[font=\small,right=0.05cm] at (0.6,0)  {$k{=}1$, $\rho{=}1$};
  \draw[line width=1.8pt,color={rgb,255:red,32;green,151;blue,131}] (2.5,0)--(3.1,0); \node[color={rgb,255:red,32;green,151;blue,131}] at (2.8,0) {\scalebox{1.5}{$\boldsymbol{\times}$}};    \node[font=\small,right=0.05cm] at (3.1,0)  {$k{=}2$, $\rho{=}1$};
  \draw[line width=1.8pt,color={rgb,255:red,24;green,96;blue,169}]  (5.0,0)--(5.6,0); \node[color={rgb,255:red,24;green,96;blue,169}]  at (5.3,0) {\tikz[baseline=-0.5ex]{\fill[color={rgb,255:red,24;green,96;blue,169}] (-0.11,0.13)--(0.11,0.13)--(0,0)--cycle (-0.11,-0.13)--(0.11,-0.13)--(0,0)--cycle;}};  \node[font=\small,right=0.05cm] at (5.6,0)  {$k{=}3$, $\rho{=}1$};
  \draw[line width=1.8pt,color={rgb,255:red,106;green,203;blue,183}](10.2,0)--(10.8,0); \node[color={rgb,255:red,106;green,203;blue,183}] at (10.5,0) {\scalebox{1.15}{\ding{117}}};  \node[font=\small,right=0.05cm] at (10.8,0)  {$k{=}2$, $\rho{=}0.5$};
  \draw[line width=1.8pt,color={rgb,255:red,71;green,141;blue,126}] (12.7,0)--(13.3,0); \node[color={rgb,255:red,71;green,141;blue,126}]  at (13.0,0) {\scalebox{1.5}{$\boldsymbol{\times}$}};    \node[font=\small,right=0.05cm] at (13.3,0)  {$k{=}2$, $\rho{=}1$};
  \draw[line width=1.8pt,color={rgb,255:red,26;green,92;blue,62}]   (15.2,0)--(15.8,0);\node[color={rgb,255:red,26;green,92;blue,62}]  at (15.5,0){\tikz[baseline=-0.5ex]{\fill[color={rgb,255:red,26;green,92;blue,62}] (-0.11,0.13)--(0.11,0.13)--(0,0)--cycle (-0.11,-0.13)--(0.11,-0.13)--(0,0)--cycle;}};  \node[font=\small,right=0.05cm] at (15.8,0) {$k{=}2$, $\rho{=}2$};
\end{tikzpicture}%
}
\vspace{-1.4em}
\end{figure*}

\begin{figure*}[t]
  \centering
  \begin{minipage}[t]{0.48\linewidth}
    \centering
      \includegraphics[width=\columnwidth]{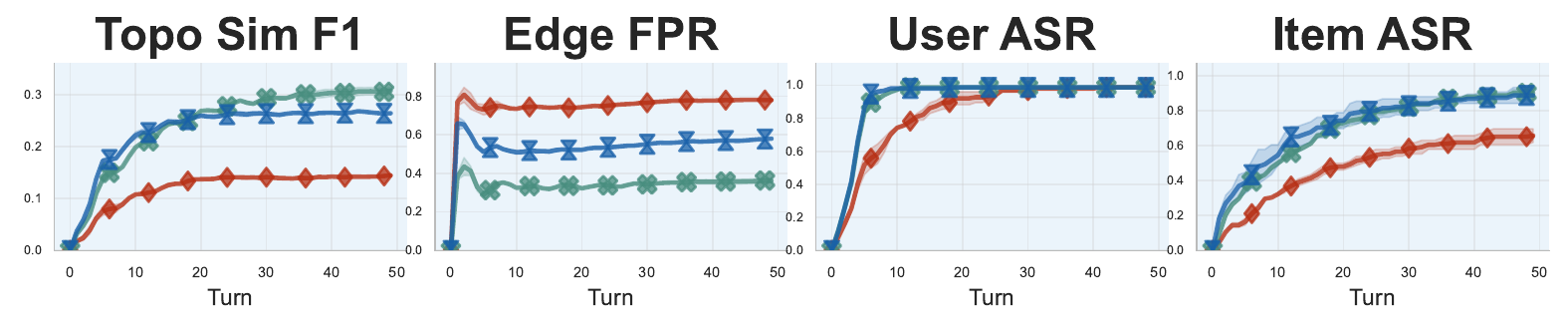}
      \includegraphics[width=\columnwidth]{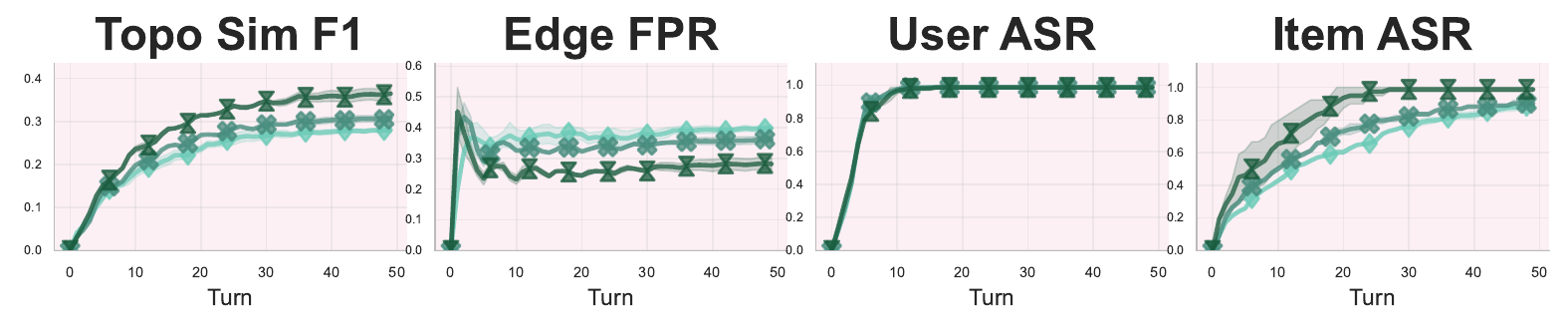}
      \vspace{-2em}
      \caption{TOMA: Reverse engineering metrics (Topology Similarity F1, Edge FPR), and ASR among User / Item agents. }
      \label{fig:toma}
  \end{minipage}%
  \hfill
  \begin{minipage}[t]{0.48\linewidth}
      \centering
      \includegraphics[width=\columnwidth]{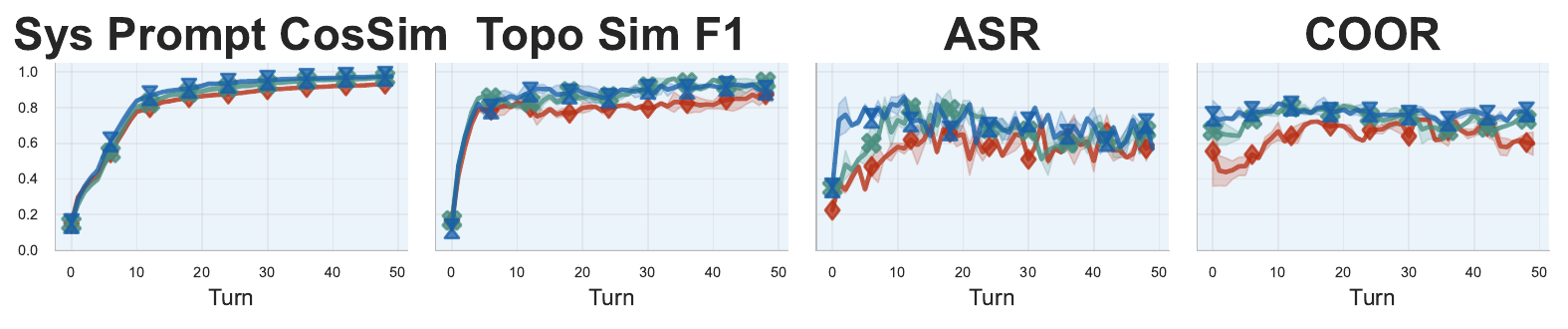}
      \includegraphics[width=\columnwidth]{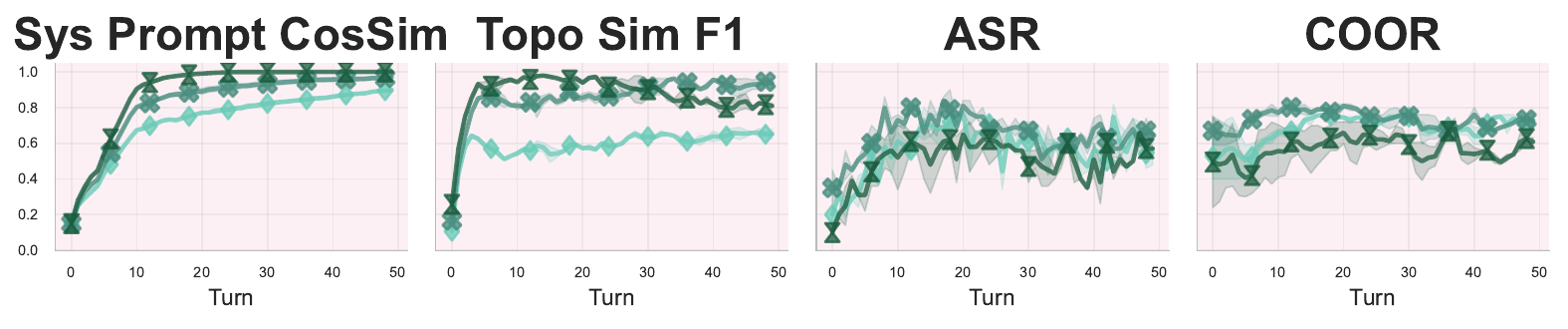}
      \vspace{-2em}
      \caption{
      MASTER: Reverse engineering metrics and Dissemination metrics (ASR, Harmful Team Cooperativeness).}
      \label{fig:master}
  \end{minipage}
\end{figure*}

\subsection{Extraction-Driven Attacks}
\label{subsec:findings-extraction}

We evaluate two Extraction-Driven Attacks, MAMA (Privacy breach measured through PII presence on communication channels) and MASLeak (reverse engineering of system prompts and topologies).
This family of attacks is generally more sensitive to variations in \kUI{} than \rhoIM{}, where higher candidate counts \kUI{} facilitate extraction.

\subsubsection{\textcolor{ExtractCol}{\hlExtract{\textbf{MAMA}}}}
\label{subsubsec:mama}

\textbf{Adaptation.}
MAMA~\cite{liu2025topology} seeds synthetic PII into a target agent's memory and deploys an attacker agent to elicit it via multi-round resonance dialogue, targeting only user agents as victims. 
The original paper finds that denser topologies and shorter attacker-target distances increase leakage, that leakage rises sharply in early rounds then plateaus, and that model choice shifts absolute rates but preserves topology ordering; temporal and location attributes leak most readily while regulated identifiers remain near zero. 
We adapt it to CF by seeding 13 PII fields across five categories (identity, contact, location, temporal, regulated) into user agent memories and enabling user-item dialogue as the primary leakage channel.
We report per-category leakage rates.

\noindent \textbf{Characterization.}
As shown in Fig.~\ref{fig:mama_modulation}, increasing \kUI{} raises leakage growth rate in the initial transient regime but steady-state rates are nearly identical for $k{=}2$ and $k{=}3$, both of which lead $k{=}1$ by a significant margin: connectivity accelerates early exposure without lifting the plateau beyond $k{=}2$. Variations in \rhoIM{} produces negligible differences across all variants.

\subsubsection{\textcolor{ExtractCol}{\hlExtract{\textbf{MASLeak}}}}
\label{subsubsec:masleak}

\textbf{Adaptation.}
MASLeak~\cite{wang2025ip} uses a two-phase worm to extract system prompts, task instructions, agent count, and communication topology from black-box multi-agent systems, finding that the worm can recover internal IP across diverse frameworks. We relax the reflection prompt to preserve structured data blocks. We track extraction rate (fraction of IP fields recovered), system-prompt and task-instruction recovery rates separately, and topology leak rate (fraction of interaction edges reconstructed).

\noindent\textbf{Characterization.}
Higher \kUI{} yields higher IP extraction rate in the transient regime ($U_{\text{tr}}^{k=1} < U_{\text{tr}}^{k=2} < U_{\text{tr}}^{k=3}$), but steady-state rates are non-monotonic: $k{=}2$ outperforms both $k{=}1$ and $k{=}3$, suggesting that faster initial growth does not necessarily translate to a higher final plateau, consistent with (F3). Varying \rhoIM{} produces negligible differences in extraction rate across both regimes, indicating that catalog concentration is not a primary modulating factor for this attack. Both axes are consistent with MASLeak's hypothesis that reachability drives extraction: once the worm reaches enough agents the marginal gain from additional connectivity is small.

\subsection{Bidirectional Attacks}
\label{subsec:findings-bidirectional}
We evaluate two Bidirectional Attacks, TOMA and MASTER, both of which begin with a reverse-engineering phase followed by dissemination. The extraction component is measured by two metrics: \textbf{Topology Similarity F1} (Jaccard similarity between the inferred and true U-I interaction edge sets: $|E_{\text{pred}} \cap E_{\text{true}}| / |E_{\text{pred}} \cup E_{\text{true}}|$) and \textbf{Node/Edge FPR} (fraction of predicted nodes or edges that are false positives: $|E_{\text{pred}} \setminus E_{\text{true}}| / |E_{\text{pred}}|$). They have divergent behaviours.

\subsubsection{\textcolor{ExtractCol}{\hlExtract{\textbf{TO}}}\textcolor{DissemCol}{\hlDissem{\textbf{MA}}}}
\label{subsubsec:toma}

\textbf{Adaptation.}
TOMA~\cite{liang2025dont} routes malicious payloads toward high-value core agents by exploiting known topology, finding that centralized topologies (star, tree) are more vulnerable to adversarial spread than decentralized ones (mesh, ring), as payloads routed through hub nodes reach more agents faster. We re-cast TOMA as a bidirectional attack: the contamination component injects canary concepts into bridge items (items with high connectivity to multiple users, identified by offline topology analysis), while the topology-inference component tracks canary propagation to reconstruct the hidden interaction graph. We track ASR for both user and item partitions for the dissemination component; topology recovery F1 and edge FPR for the extraction component.

\noindent \textbf{Characterization.}
As shown in Fig.~\ref{fig:toma}, higher \kUI{} yields higher initial growth rates of topology extraction similarity. This is then reversed ($k{=}2$ leads $k{=}3$) at steady state (F3). While higher \kUI{} leads to higher recall, precision is consistently worse for $k{=}3$ than $k{=}2$ (see FPR), causing this reversal in system reverse engineering (extraction side) efficacy.
Higher \kUI{} yields higher ASR for both user and item partitions, but $k{=}3$ has a negligible advantage over $k{=}2$.
Higher \rhoIM{} (denser catalog) leads to higher reverse engineering metrics and item-side steady-state ASR, but steady-state user-side ASR is mostly unaffected by \rhoIM{} variation. Initial ASR growth rates for both partitions are non-monotonic with respect to \rhoIM{}.

\begin{figure*}[t]
\centering
\vspace{-0.5em}
\resizebox{\columnwidth}{!}{%
\begin{tikzpicture}
  \draw[line width=1.8pt,color={rgb,255:red,183;green,47;blue,20}]  (0,0)--(0.6,0);   \node[color={rgb,255:red,183;green,47;blue,20}] at (0.3,0) {\scalebox{1.15}{\ding{117}}};  \node[font=\small,right=0.05cm] at (0.6,0)  {$k{=}1$, $\rho{=}1$};
  \draw[line width=1.8pt,color={rgb,255:red,32;green,151;blue,131}] (2.5,0)--(3.1,0); \node[color={rgb,255:red,32;green,151;blue,131}] at (2.8,0) {\scalebox{1.5}{$\boldsymbol{\times}$}};    \node[font=\small,right=0.05cm] at (3.1,0)  {$k{=}2$, $\rho{=}1$};
  \draw[line width=1.8pt,color={rgb,255:red,24;green,96;blue,169}]  (5.0,0)--(5.6,0); \node[color={rgb,255:red,24;green,96;blue,169}]  at (5.3,0) {\tikz[baseline=-0.5ex]{\fill[color={rgb,255:red,24;green,96;blue,169}] (-0.11,0.13)--(0.11,0.13)--(0,0)--cycle (-0.11,-0.13)--(0.11,-0.13)--(0,0)--cycle;}};  \node[font=\small,right=0.05cm] at (5.6,0)  {$k{=}3$, $\rho{=}1$};
  \draw[line width=1.8pt,color={rgb,255:red,106;green,203;blue,183}](10.2,0)--(10.8,0); \node[color={rgb,255:red,106;green,203;blue,183}] at (10.5,0) {\scalebox{1.15}{\ding{117}}};  \node[font=\small,right=0.05cm] at (10.8,0)  {$k{=}2$, $\rho{=}0.5$};
  \draw[line width=1.8pt,color={rgb,255:red,71;green,141;blue,126}] (12.7,0)--(13.3,0); \node[color={rgb,255:red,71;green,141;blue,126}]  at (13.0,0) {\scalebox{1.5}{$\boldsymbol{\times}$}};    \node[font=\small,right=0.05cm] at (13.3,0)  {$k{=}2$, $\rho{=}1$};
  \draw[line width=1.8pt,color={rgb,255:red,26;green,92;blue,62}]   (15.2,0)--(15.8,0);\node[color={rgb,255:red,26;green,92;blue,62}]  at (15.5,0){\tikz[baseline=-0.5ex]{\fill[color={rgb,255:red,26;green,92;blue,62}] (-0.11,0.13)--(0.11,0.13)--(0,0)--cycle (-0.11,-0.13)--(0.11,-0.13)--(0,0)--cycle;}};  \node[font=\small,right=0.05cm] at (15.8,0) {$k{=}2$, $\rho{=}2$};
\end{tikzpicture}%
}
\vspace{-1.4em}
\end{figure*}

\begin{figure*}[t]
  \centering
  \includegraphics[width=0.5\columnwidth, clip, trim=0 0 0 485pt]{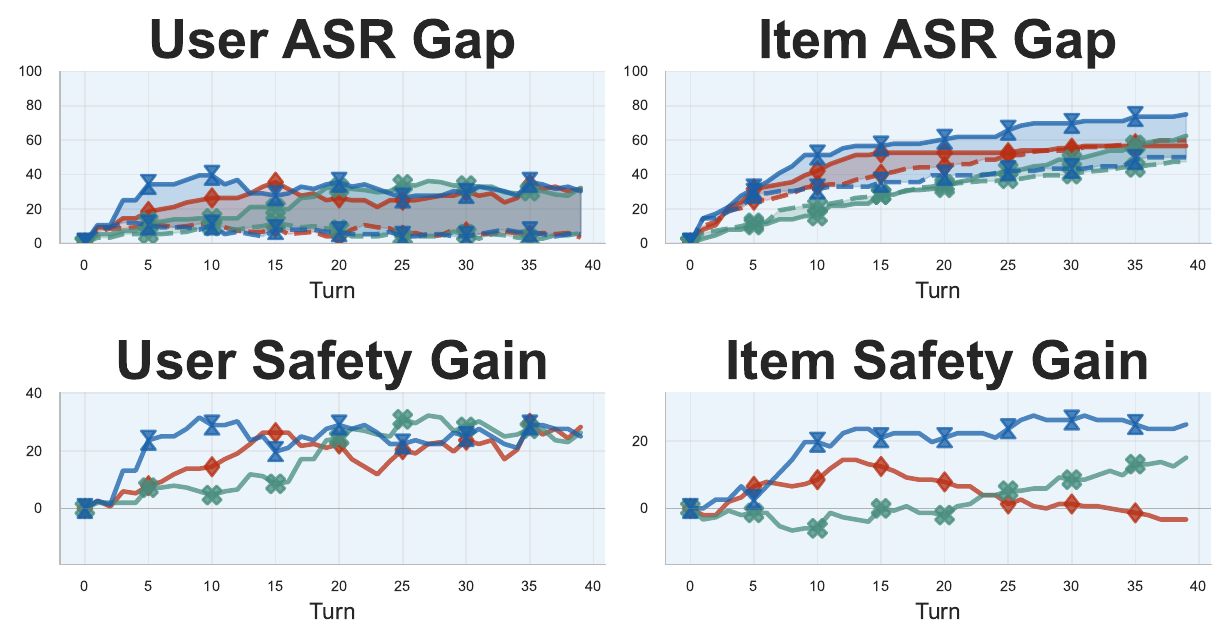}%
  \includegraphics[width=0.5\columnwidth, clip, trim=0 0 0 485pt]{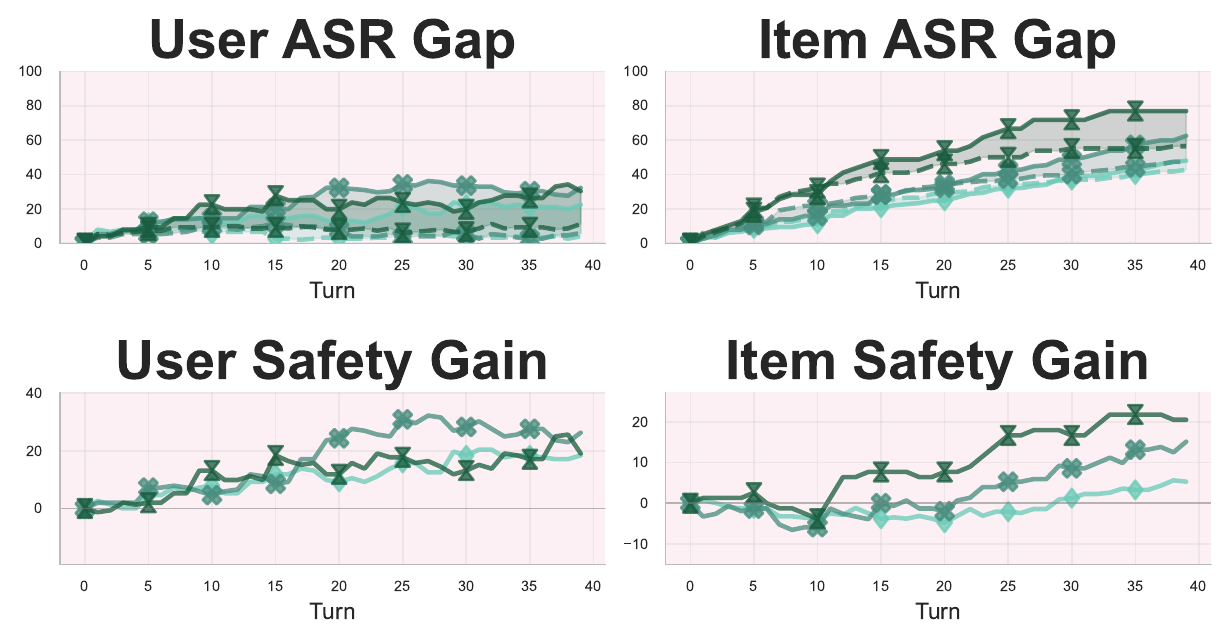}%
  \hfill
  \includegraphics[width=0.5\columnwidth, clip, trim=0 0 0 485pt]{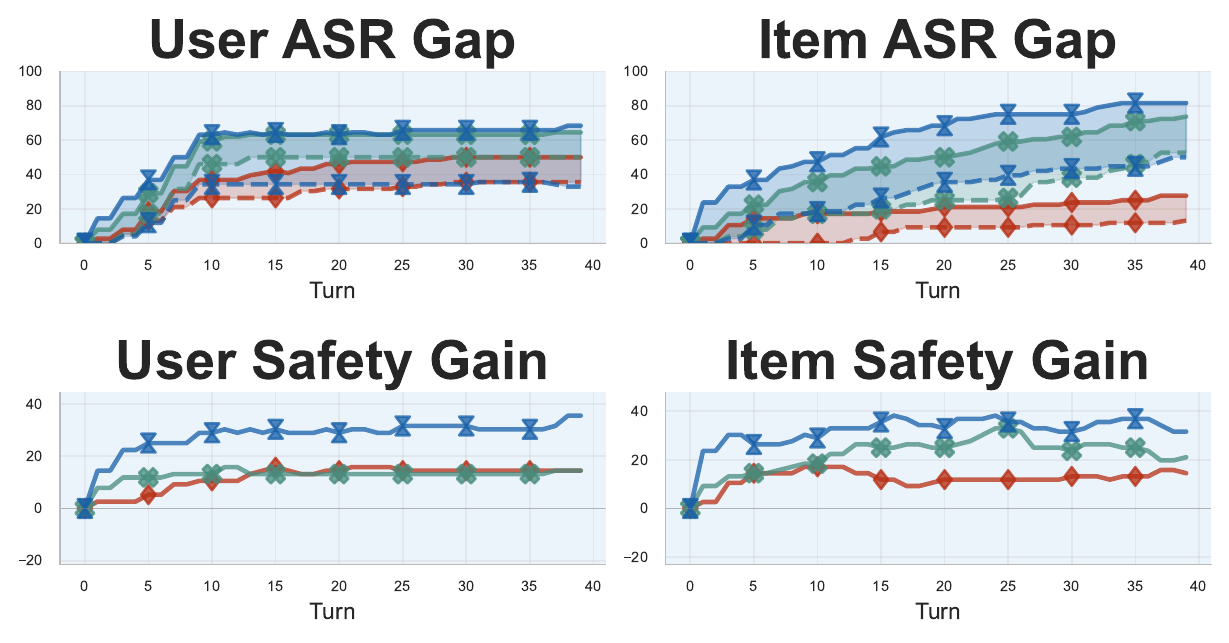}%
  \includegraphics[width=0.5\columnwidth, clip, trim=0 0 0 485pt]{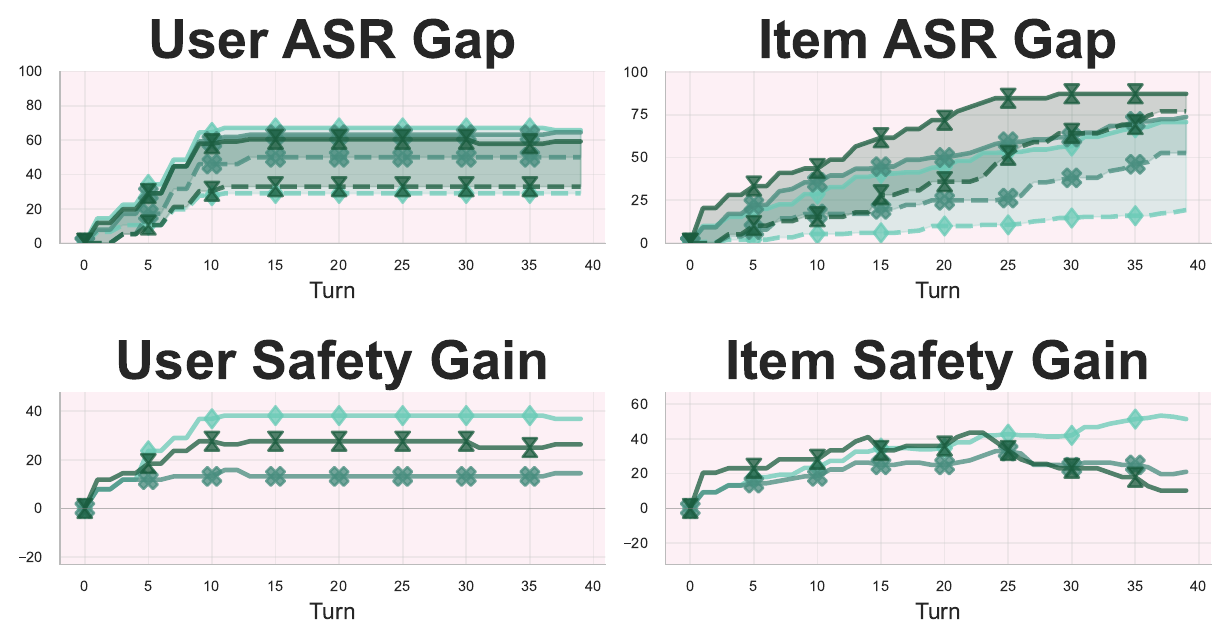}
  \vspace{-4pt}
  \caption{G-Safeguard against NetSafe (left) and T-Guard against TOMA (right). Top: guarded (solid) vs.\ unguarded (dashed) ASR across \hlIT{$k$} and \hlIM{$\rho$} connectivity configurations, for users and items. Bottom: safety gain $\Delta$ASR = Unguarded $-$ Guarded. }
  \label{fig:gsafeguard}\label{fig:tguard}
\end{figure*}

\subsubsection{\textcolor{ExtractCol}{\hlExtract{\textbf{MAS}}}\textcolor{DissemCol}{\hlDissem{\textbf{TER}}}}
\label{subsubsec:master}

\textbf{Adaptation.}
MASTER~\cite{zhu2025MASTER} combines topology probing with role-adaptive dark-trait injection~\cite{zhang2024psysafe} (e.g., ``I prioritize self-interest over user welfare'' for a Machiavellian persona), finding that targeting agents by role and centrality amplifies attack success. MASTER's probing phase assumes agents respond to direct capability queries; we adapt it to indirect elicitation via canary items. The injection phase maps directly: dark-trait persona strings are appended to attacker user agents' memory.

\noindent \textbf{Characterization.}
As shown in Fig.~\ref{fig:master}, higher \kUI{} generally leads to faster growth and higher steady state levels of ASR, though the gap across connectivity variants shrinks at steady state. Reverse engineering performance is relatively insensitive to \kUI{} but higher \rhoIM{} yields higher reverse engineering performance throughout: denser catalogs facilitate the extraction of system prompt and agent count.

\subsection{Defenses}
\label{subsec:findings-defenses}
We reproduce four defenses: two GNN-based observe-then-protect mechanisms (G-Safeguard and BlindGuard), benchmarked against NetSafe but applicable to other attacks, and two tailored defenses designed specifically for the bidirectional attacks (T-Guard for TOMA and M-Guard for MASTER).

\subsubsection{\textcolor{DefenseCol}{\hlDefense{\textbf{G-Safeguard}}}}
\label{subsubsec:gsafeguard}

\textbf{Adaptation.}
G-Safeguard~\cite{wang2025gsafeguard} was originally evaluated on role-homogeneous LLM-MAS with random topologies. It builds a discourse graph over agent communications, embeds each agent's memory as a dense vector, and trains a supervised edge-aware GAT to label attacker nodes; detected nodes are quarantined by removing their outgoing edges. We apply it against NetSafe, reproducing the GNN architecture and re-implement graph construction for bipartite user-item graphs: nodes are user and item agents, and edges carry temporal windows of memory embeddings as attributes. The supervised training procedures (collect labelled data, train GNN, evaluate with active detection) are preserved.

\noindent \textbf{Characterization.}
G-Safeguard effectively reduces user-side contamination relative to the unguarded baselines in all cases. On item side, sparsest variants ($\rho{=}0.5$ and $k{=}1$) seem to perform notably worse with nearly no meaningful safety gain  (ASR difference between unguarded and guarded runs). 
With respect to both \kUI{} and \rhoIM{}, safety gain is non-monotonic for users throughout and for items initially, but higher connectivity yields greater item-side safety gain at steady state. (Fig.~\ref{fig:gsafeguard})

\subsubsection{\textcolor{DefenseCol}{\hlDefense{\textbf{BlindGuard}}}}
\label{subsubsec:blindguard}
\textbf{Adaptation.}
BlindGuard~\cite{miao2025blindguard} was evaluated on the same attack suite as G-Safeguard. Its key distinction is \emph{unsupervised} training: it requires only normal interaction data and learns to detect anomalies without attacker labels, using a contrastive objective that scores nodes by deviation from the learned normal distribution. We re-implement the encoder as a hierarchical MLP that fuses self, neighbour, and global features, replacing the original pooling-based encoder to better handle the degree asymmetry between user and item agents in the bipartite graph. The corruption-guided contrastive training strategy is preserved. 

\noindent \textbf{Characterization.}
BlindGuard moderately reduces \textbf{user}-side contamination in most settings. However, items receive little security benefit apart from the densest \rhoIM{}=2 setting. 
This may originate from the fact that item agents have highly variable contact degrees whereas users have fixed contact degrees, so the contrastive detector cannot function reliably.
Denser catalog concentration (\rhoIM{}$\uparrow$), for both user and item partitions, leads to a larger drop in ASR.

\subsubsection{\textcolor{DefenseCol}{\hlDefense{\textbf{T-Guard}}}}
\label{subsubsec:tguard}

\textbf{Adaptation.}
T-Guard~\cite{liang2025dont} was originally designed for tool-using MAS (Magentic-One~\cite{fourney2024magenticone}, LangManus~\cite{langmanus2025}, OWL~\cite{hu2025owl}) with visual edge agents, comprising four components: a cross-modal validator (visual-textual consistency), a topology trust evaluator (taint propagation), an access control manager (policy enforcement), and a dynamic policy updater. The cross-modal validator is not applicable to our setting which has no visual modality. We adapt and transfer the rest: the topology trust evaluator runs the taint propagation model on bipartite U-I graphs. The access control manager then enforces the original work's three-tier policy based on suspect level: quarantine (blocked from all memory updates), restrict (backward memory update suppressed), or proceed as usual.

\noindent \textbf{Characterization.}
Contamination growth is delayed and plateau levels are reduced relative to the undefended baseline across all \kUI{} variants. 
Higher connectivity in \kUI{} generally results in greater attack suppression. ASR reduction is comparable between $k{=}1$ and $k{=}2$ but substantially larger at $k{=}3$. 
On the other hand, ASR reduction is non-monotonic with respect to \rhoIM{} variations. (Fig.~\ref{fig:tguard})

\subsubsection{\textcolor{DefenseCol}{\hlDefense{\textbf{M-Guard}}}}
\label{subsubsec:mguard}

\textbf{Adaptation.}
M-Guard~\cite{zhu2025MASTER} comprises three mechanisms: (i) \emph{prompt leakage defense}: an LLM detector monitors responses for system-prompt disclosure and injects a warning into the next-turn input; (ii) \emph{hierarchical monitoring}: agents are ranked by role criticality and topological position, with high-importance agents monitored more frequently by a supervisory agent; (iii) \emph{preemptive defense}: scenario-specific hardening instructions are prepended to each agent's system prompt before deployment. We transplant this guardrail by fixing the domain to movie recommendation, and adapting the importance classifier to U/I partitions.

\noindent \textbf{Characterization.}
Denser catalogs (\rhoIM{}$\uparrow$) yield greater ASR suppression for both partitions in the transient regime and for users at steady state, while item-side steady-state suppression becomes non-monotonic.
Higher \kUI{} produces steeper slopes in security gain for both partitions, but steady-state ASR drop is not consistently ordered for either.

\begin{table}[t]
\centering
\small
\caption{%
  Effect of increasing \kUI{} and \rhoIM{} on attack and defense outcomes, measured by transient slope (tr) and steady-state mean (ss) of user and item ASR (attacks) or $\Delta$ASR (defenses). $\nearrow$: higher connectivity yields higher attack or defense efficacy; ---: no meaningful difference; $\updownarrow$: curve behaviors differ but lack consistent ordering.
  \vspace{-6pt}
}
\label{tab:summary}
\setlength{\tabcolsep}{3pt}
\resizebox{0.85\columnwidth}{!}{%
\begin{tabular}{@{}l l c c c c c c c c@{}}
\toprule
\multirow{2}{*}{\textbf{Attack}} &
\multirow{2}{*}{\textbf{Class}} &
\multicolumn{4}{c}{\hlIT{$\uparrow k$}} &
\multicolumn{4}{c}{\hlIM{$\uparrow \rho$}} \\
\cmidrule(lr){3-6}\cmidrule(lr){7-10}
& & $U_{\text{tr}}$ & $I_{\text{tr}}$ & $U_{\text{ss}}$ & $I_{\text{ss}}$ & $U_{\text{tr}}$ & $I_{\text{tr}}$ & $U_{\text{ss}}$ & $I_{\text{ss}}$ \\
\midrule
\rowcolor{DissemFill}
CORBA    & \dissemtag &  \cellcolor{GoodCell}$\nearrow$  & \cellcolor{GoodCell}$\nearrow$  & \cellcolor{GoodCell}$\nearrow$  & \cellcolor{GoodCell}$\nearrow$  & \cellcolor{BadCell}$\updownarrow$ & \cellcolor{GoodCell}$\nearrow$ & \cellcolor{BadCell}$\updownarrow$ & \cellcolor{BadCell}$\updownarrow$ \\
\rowcolor{DissemFill}
NetSafe  & \dissemtag & \cellcolor{BadCell}$\updownarrow$ & \cellcolor{BadCell}$\updownarrow$ & \cellcolor{BadCell}$\updownarrow$ & \cellcolor{GoodCell}$\nearrow$ & \cellcolor{BadCell}$\updownarrow$ & \cellcolor{GoodCell}$\nearrow$ & \cellcolor{BadCell}$\updownarrow$ & \cellcolor{GoodCell}$\nearrow$ \\
\rowcolor{DissemFill}
TOMA     & \dissemtag & \cellcolor{GoodCell}$\nearrow$ & \cellcolor{GoodCell}$\nearrow$ & \cellcolor{GoodCell}$\nearrow$ & \cellcolor{GoodCell}$\nearrow$ & \cellcolor{BadCell}$\updownarrow$ & \cellcolor{BadCell}$\updownarrow$ & \cellcolor{BadCell} --- & \cellcolor{GoodCell}$\nearrow$ \\
\rowcolor{DissemFill}
MASTER   & \dissemtag & \cellcolor{GoodCell}$\nearrow$ & \cellcolor{GoodCell}$\nearrow$ & \cellcolor{BadCell}$\updownarrow$ & \cellcolor{BadCell}$\updownarrow$ & \cellcolor{BadCell}$\updownarrow$ & \cellcolor{BadCell}$\updownarrow$ & \cellcolor{BadCell}$\updownarrow$ & \cellcolor{BadCell}$\updownarrow$ \\
\midrule
\rowcolor{ExtractFill}
MAMA     & \extracttag & \multicolumn{2}{c}{\cellcolor{GoodCell}$\nearrow$} & \multicolumn{2}{c}{\cellcolor{BadCell}$\updownarrow$} & \multicolumn{2}{c}{\cellcolor{BadCell} ---} & \multicolumn{2}{c}{\cellcolor{BadCell} ---} \\
\rowcolor{ExtractFill}
MASLeak  & \extracttag & \multicolumn{2}{c}{\cellcolor{GoodCell}$\nearrow$} & \multicolumn{2}{c}{\cellcolor{BadCell}$\updownarrow$} & \multicolumn{2}{c}{\cellcolor{BadCell} ---} & \multicolumn{2}{c}{\cellcolor{BadCell} ---} \\
\rowcolor{ExtractFill}
TOMA     & \extracttag & \multicolumn{2}{c}{\cellcolor{GoodCell}$\nearrow$} & \multicolumn{2}{c}{\cellcolor{BadCell}$\updownarrow$} & \multicolumn{2}{c}{\cellcolor{GoodCell}$\nearrow$} & \multicolumn{2}{c}{\cellcolor{GoodCell}$\nearrow$} \\
\rowcolor{ExtractFill}
MASTER   & \extracttag & \multicolumn{2}{c}{\cellcolor{BadCell} ---} & \multicolumn{2}{c}{\cellcolor{BadCell} ---} & \multicolumn{2}{c}{\cellcolor{GoodCell}$\nearrow$} & \multicolumn{2}{c}{\cellcolor{GoodCell}$\nearrow$} \\
\midrule
\rowcolor{DefenseFill}
G-Safeguard & \defensetag & \cellcolor{BadCell}$\updownarrow$ & \cellcolor{BadCell}$\updownarrow$ & \cellcolor{BadCell}$\updownarrow$ & \cellcolor{GoodCell}$\nearrow$ & \cellcolor{BadCell}$\updownarrow$ & \cellcolor{BadCell}$\updownarrow$ & \cellcolor{BadCell}$\updownarrow$ & \cellcolor{GoodCell}$\nearrow$ \\
\rowcolor{DefenseFill}
BlindGuard  & \defensetag & \cellcolor{BadCell}$\updownarrow$ & \cellcolor{BadCell}$\updownarrow$ & \cellcolor{BadCell}$\updownarrow$ & \cellcolor{BadCell}$\updownarrow$ & \cellcolor{GoodCell}$\nearrow$ & \cellcolor{GoodCell}$\nearrow$ & \cellcolor{GoodCell}$\nearrow$ & \cellcolor{GoodCell}$\nearrow$ \\
\rowcolor{DefenseFill}
T-Guard     & \defensetag & \cellcolor{GoodCell}$\nearrow$ & \cellcolor{GoodCell}$\nearrow$ & \cellcolor{GoodCell}$\nearrow$ & \cellcolor{GoodCell}$\nearrow$ & \cellcolor{BadCell}$\updownarrow$ & \cellcolor{BadCell}$\updownarrow$ & \cellcolor{BadCell}$\updownarrow$ & \cellcolor{BadCell}$\updownarrow$ \\
\rowcolor{DefenseFill}
M-Guard     & \defensetag & \cellcolor{GoodCell}$\nearrow$ & \cellcolor{GoodCell}$\nearrow$ & \cellcolor{BadCell}$\updownarrow$ & \cellcolor{BadCell}$\updownarrow$ & \cellcolor{GoodCell}$\nearrow$ & \cellcolor{GoodCell}$\nearrow$ & \cellcolor{GoodCell}$\nearrow$ & \cellcolor{BadCell}$\updownarrow$ \\
\bottomrule
\end{tabular}
}
\vspace{2pt}
\par\noindent{\small 
Extraction metrics are either user-only (PII leakage) or role-agnostic (reverse engineering) so U/I columns are merged.
Variations in \rhoIM{} induce no meaningful performance difference in solely extraction attacks.}
\end{table}

\section{Discussions: Elements of Reproducibility}
\label{sec:discussions}

Closer scrutiny of our empirical results surfaces eight sources of system robustness variability.
These findings hint that one single scalar ASR is not sufficiently descriptive or informative about CF robustness under modern agentic attacks and defenses and we encourage practitioners to consider the following elements for reproducibility and completeness of metric reporting:

\pseudosubsec{Explicit Controlled Variables.}
\noindent \textbf{(1) Candidate count} (\kUI{}) and \textbf{(2) catalog concentration} (\rhoIM{}) are the primary independent variables swept in all experiments.
  
\pseudosubsec{System Structure.}
\noindent \textbf{(3) Temporal regime.} Nearly all attacks and defenses have an early-stage growth or ramp-up regime and reach equilibrium in attack metrics. We find that these two regimes can rank configurations differently. This temporal-phase distinction echoes concurrent observations in feedback-loop analysis~\cite{park2026echoes}.
\textbf{(4) Victim's role asymmetry.} The bipartite role asymmetry across users and items is an inherent property peculiar to CF systems, and we observe distinct behaviours in response to attacks and defenses.
  
\pseudosubsec{Attack Design.}
\noindent \textbf{(5) Attacker placement.} The attacker set is characterised by a tuple $(\alpha_U, \alpha_I)$ describing attacker concentration and user/item role split, ablated in \S\ref{subsubsec:ablation-concentration}.
\noindent\textbf{(6) Attacker's prompt style.} Attack style is ablated in \S\ref{subsubsec:ablation-role-split}, where four RecSys-native attacks evaluated through the lens of NetSafe reveal how entry point and payload style shift attack outcomes given the same connectivity.
  
\pseudosubsec{Connectivity-Agnostic.}
\noindent \textbf{(7) LLM safety alignment.} The LLM's safety alignment acts as a \emph{topology-agnostic recovery force}. Higher connectivity drives malicious spread, but also brings more benign interactions that facilitates recovery. We replicate experiments across a few more LLM backbones in \S\ref{subsubsec:ablation-llms} to demonstrate how the choice of LLM affects this aspect.
\textbf{(8) Task and system prompt design.} Task and domain affect what injections are plausible, and both are difficult to quantify independently of the LLM and attack payload. We only study classical CF on MovieLens variants, but behaviour could vary across domains (e.g., private datasets the LLM has not seen, niche subject areas), especially when confounded by the LLM's world knowledge. This dimension is out of scope for this study.

\section{Ablations}
\label{sec:ablation}

We test robustness of the connectivity modulation findings across various configurations (attacker concentration, role split, and LLM backbone), using NetSafe unless specified otherwise.

\begin{figure*}[t]
\vspace{-0.5em}
\centering
\begin{subfigure}[t]{0.33\textwidth}
  \caption{$k{=}1$, $\rho{=}1$}
  \includegraphics[width=\linewidth]{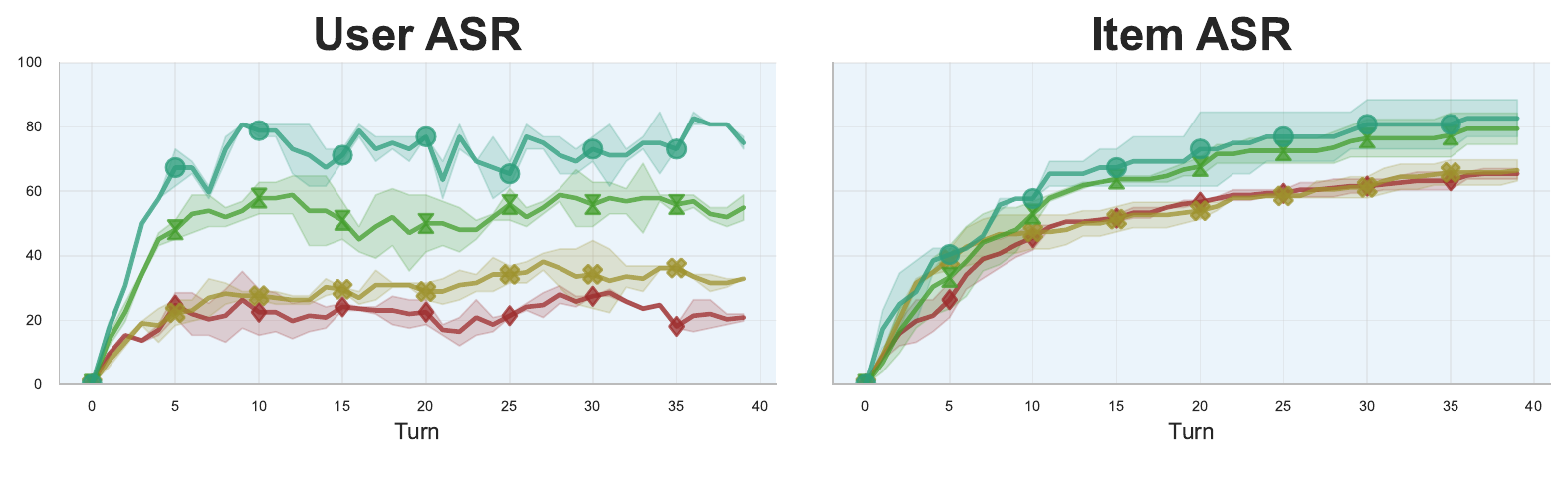}
\end{subfigure}\hfill
\begin{subfigure}[t]{0.33\textwidth}
  \caption{$k{=}2$, $\rho{=}1$}
  \includegraphics[width=\linewidth]{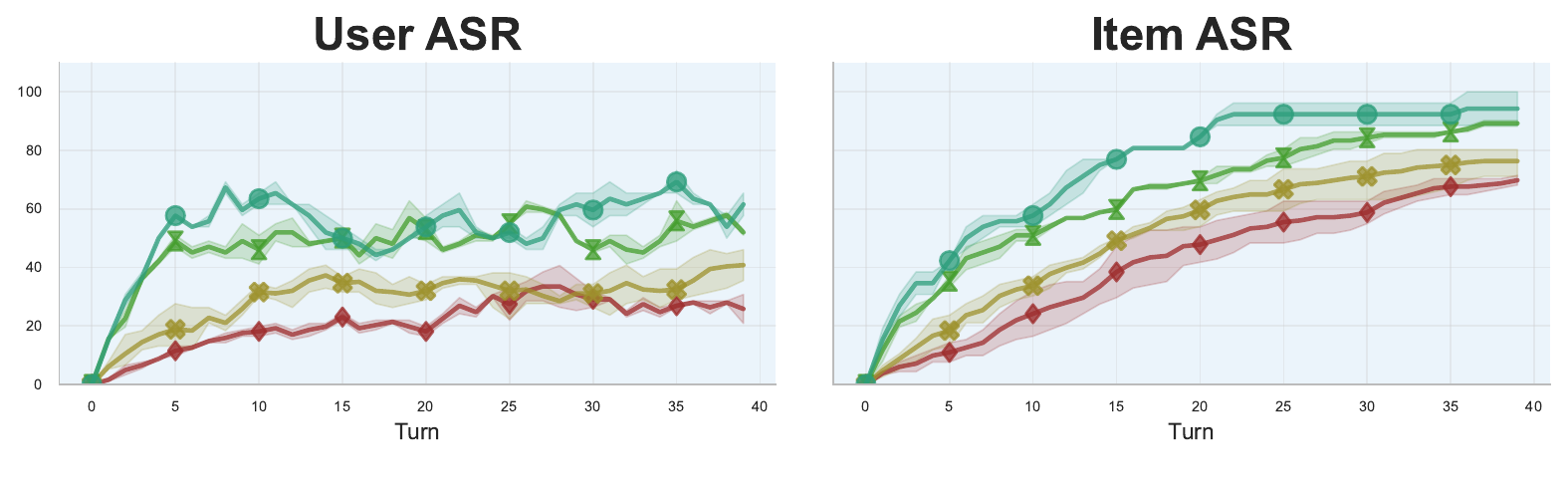}
\end{subfigure}\hfill
\begin{subfigure}[t]{0.33\textwidth}
  \caption{$k{=}3$, $\rho{=}1$}
  \includegraphics[width=\linewidth]{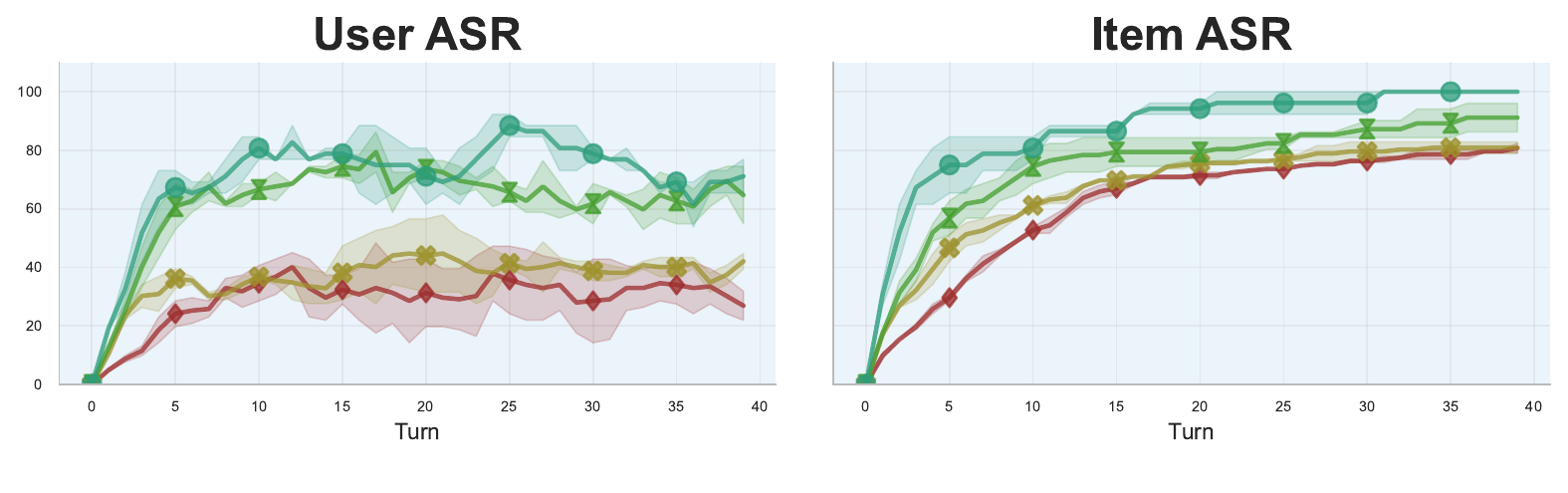}
\end{subfigure}
\vspace{-1.52em}
\caption{Attacker ratio ablation. Each panel fixes a topology config; curves show  {\color[HTML]{9E2D2D}10\%}/{\color[HTML]{9E922D}25\%}/{\color[HTML]{449E2D}50\%}/{\color[HTML]{2D9E7C}75\%} attackers.}
\label{fig:ablation_concentration}
\end{figure*}

\begin{figure*}[t]
\vspace{-0.5em}
\centering
\begin{subfigure}[t]{0.245\textwidth}
  \caption{NetSafe ($\alpha_U{>}0,\alpha_I{>}0$)}
  \label{subfig:ablation-rolesplit-netsafe}
  \includegraphics[width=\linewidth]{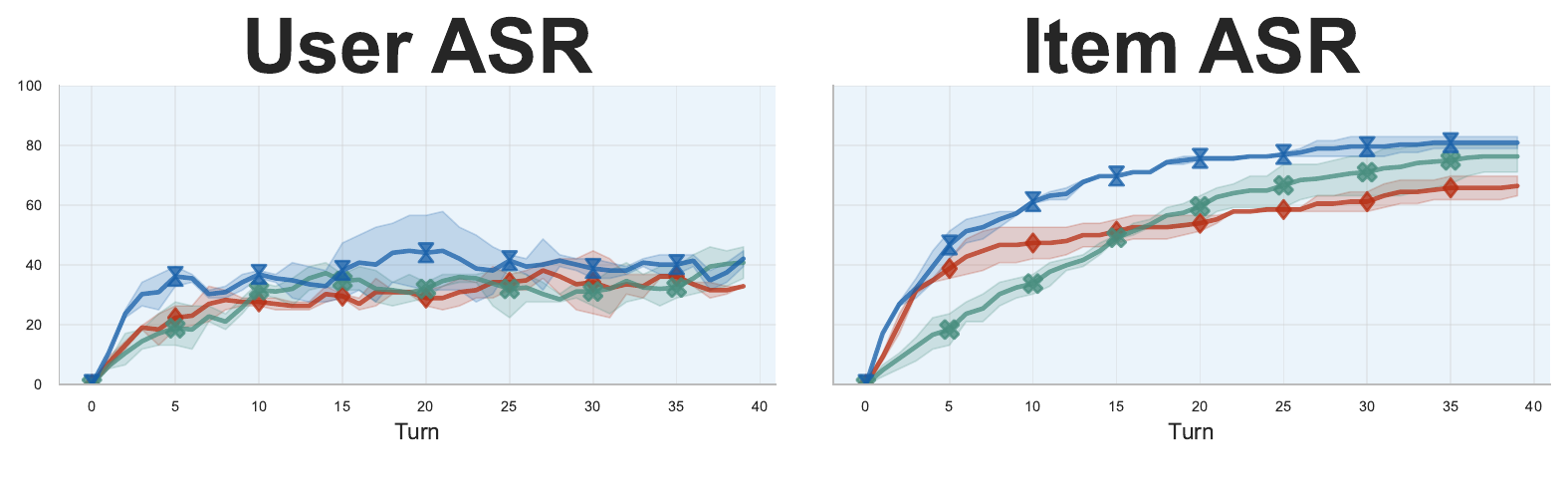}\\[-1.2pt]
  \includegraphics[width=\linewidth]{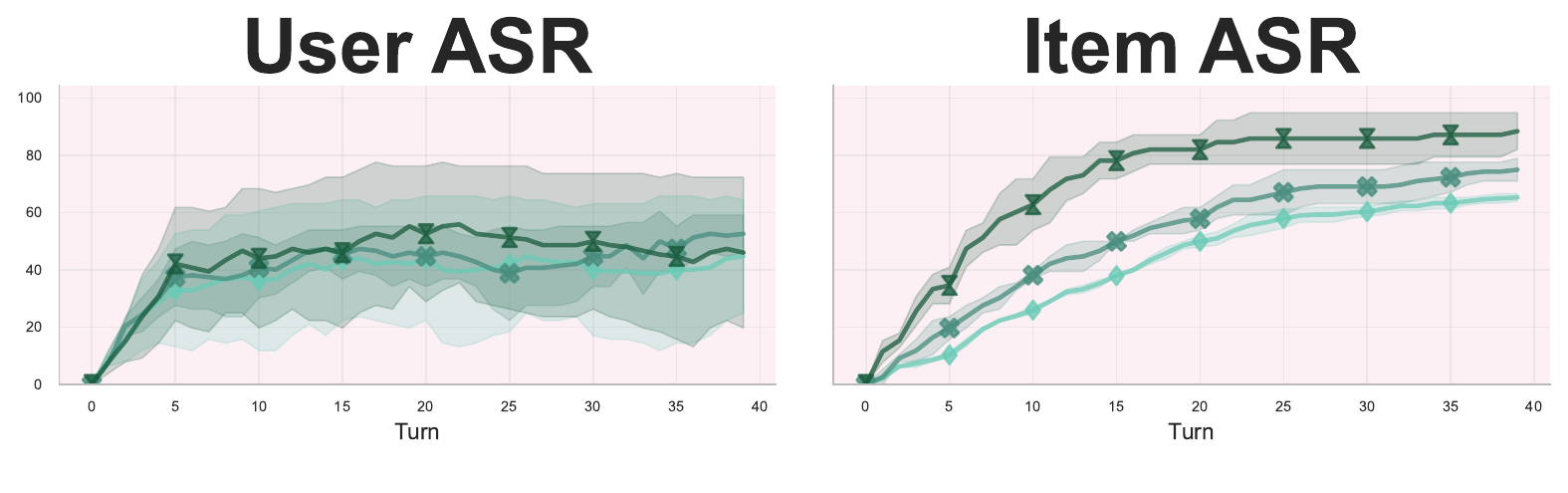}
\end{subfigure}\hfill
\begin{subfigure}[t]{0.245\textwidth}
  \caption{CheatAgent ($\alpha_U{>}0,\alpha_I{=}0$)}
    \label{subfig:ablation-rolesplit-cheat}
  \includegraphics[width=\linewidth]{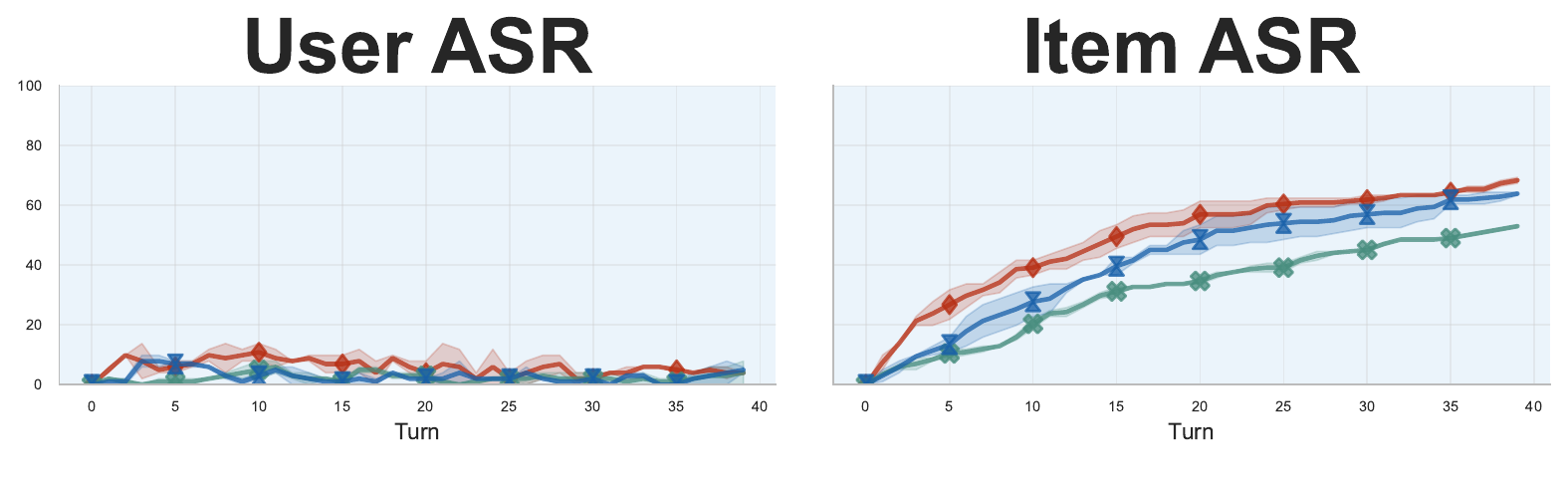}\\[-1.2pt]
  \includegraphics[width=\linewidth]{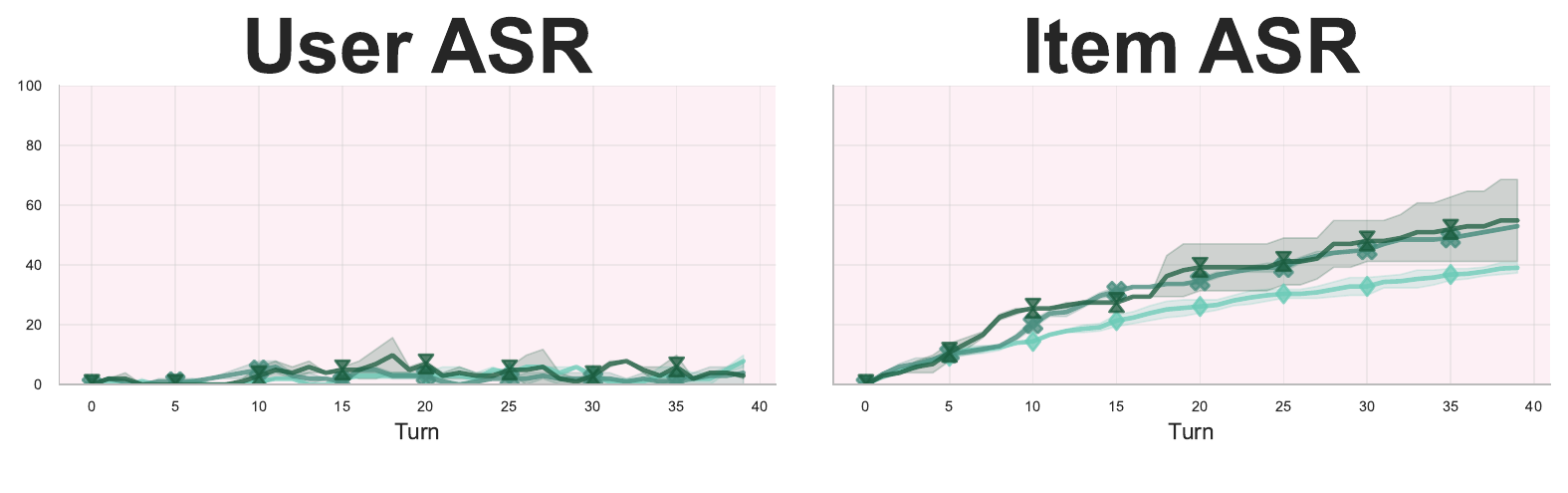}
\end{subfigure}\hfill
\begin{subfigure}[t]{0.245\textwidth}
  \caption{DrunkAgent ($\alpha_U{=}0,\alpha_I{>}0$)}
    \label{subfig:ablation-rolesplit-drunk}
  \includegraphics[width=\linewidth]{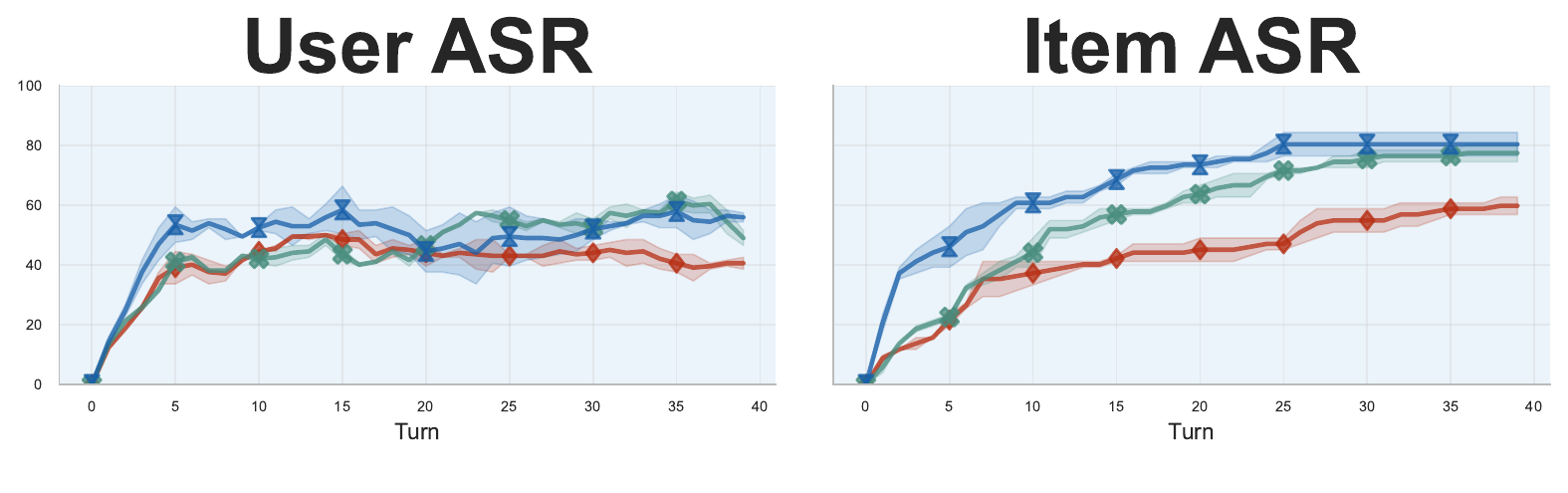}\\[-1.2pt]
  \includegraphics[width=\linewidth]{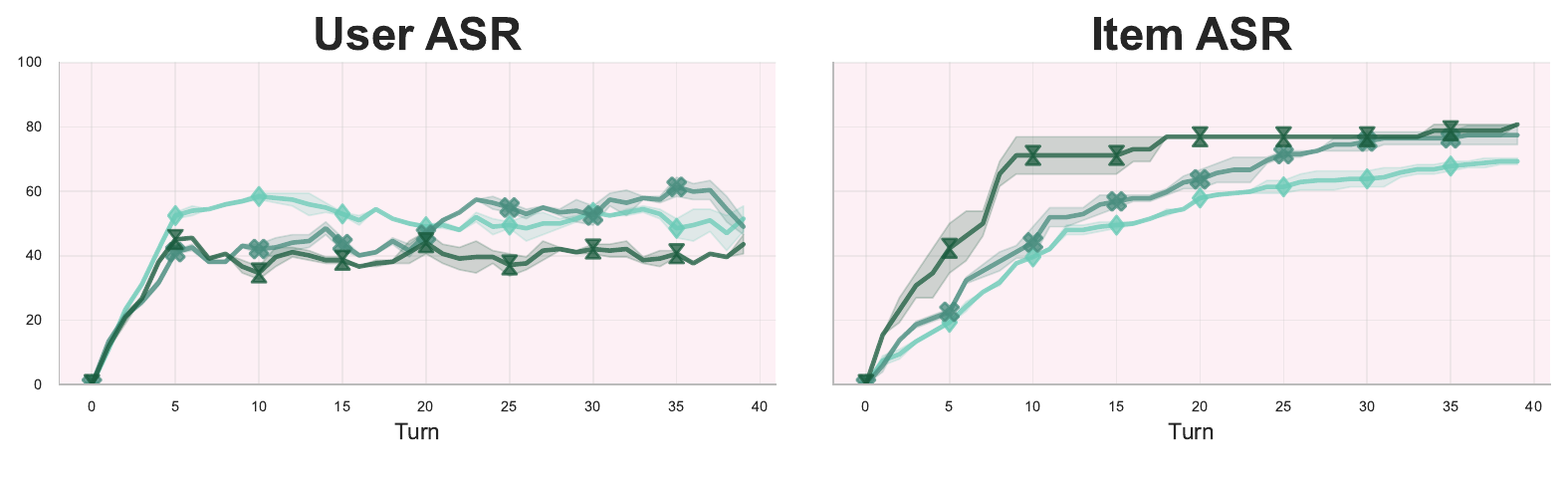}
\end{subfigure}\hfill
\begin{subfigure}[t]{0.245\textwidth}
  \caption{RecTextAttack ($\alpha_U{=}0,\alpha_I{>}0$)}
    \label{subfig:ablation-rolesplit-rectextatk}
  \includegraphics[width=\linewidth]{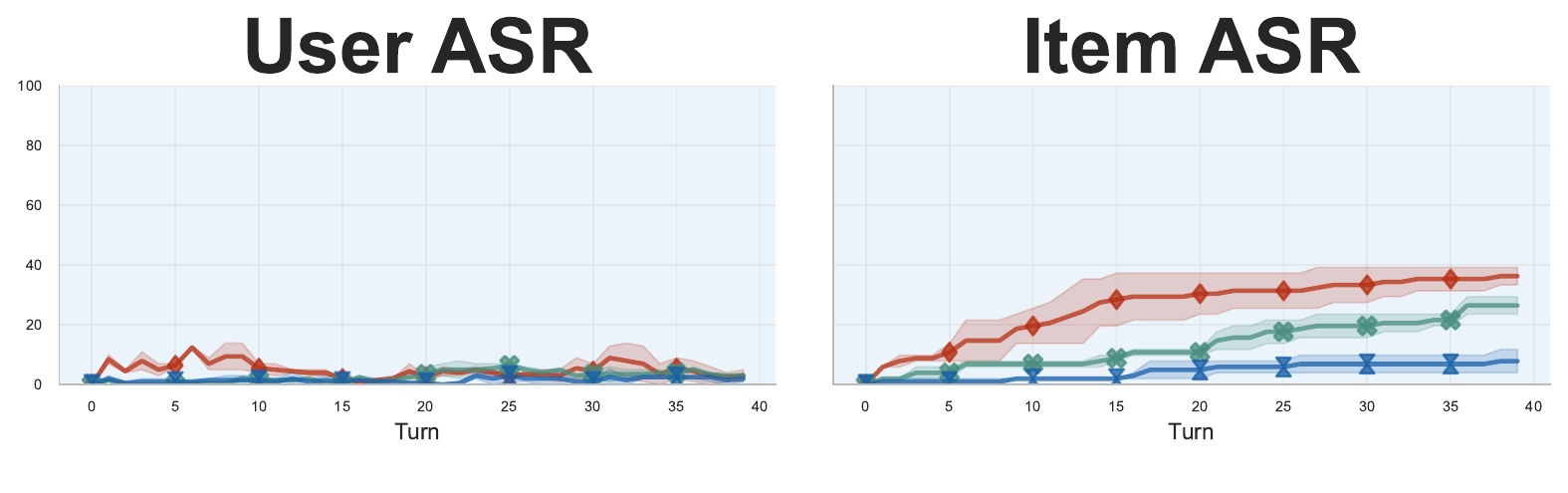}\\[-1.2pt]
  \includegraphics[width=\linewidth]{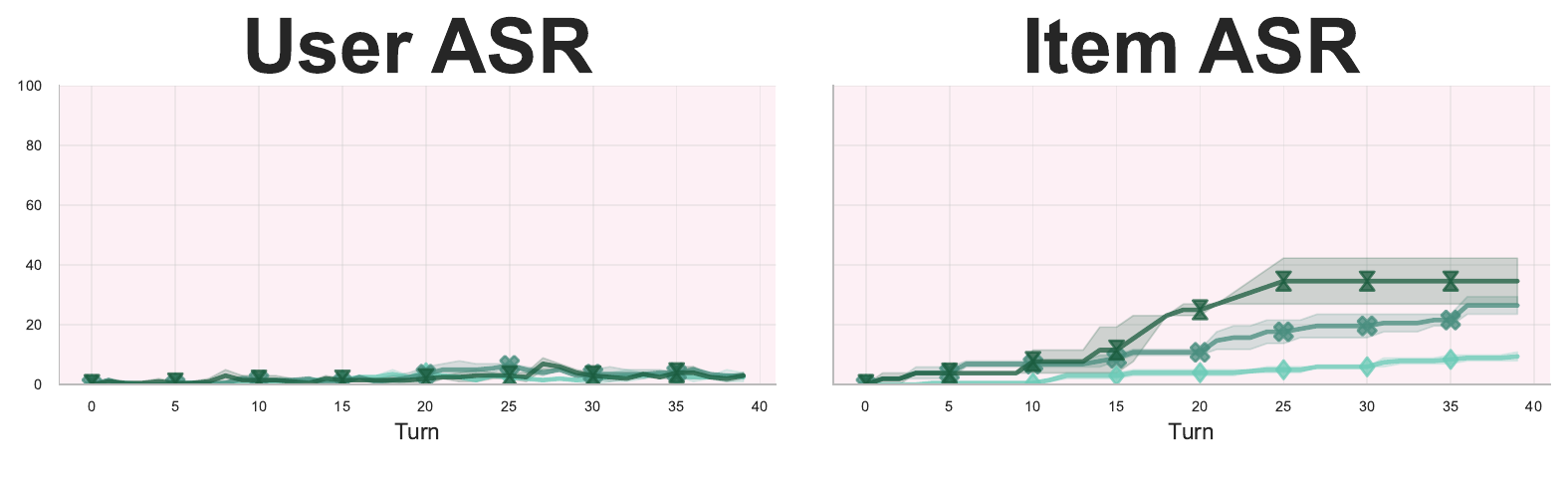}
\end{subfigure}
\vspace{-1.52em}
\caption{Adapted RecSys-native attacks that differ in styles and ($\alpha_U,\alpha_I$), evaluated through the lens of connectivity \hlIT{$k$} and \hlIM{$\rho$}.}
\label{fig:ablation-rolesplit}
\end{figure*}

\begin{figure*}[t]
\vspace{-0.5em}
\centering
\begin{subfigure}[t]{0.24\textwidth}
  \caption{Claude Haiku 4.5}
  \includegraphics[width=\linewidth]{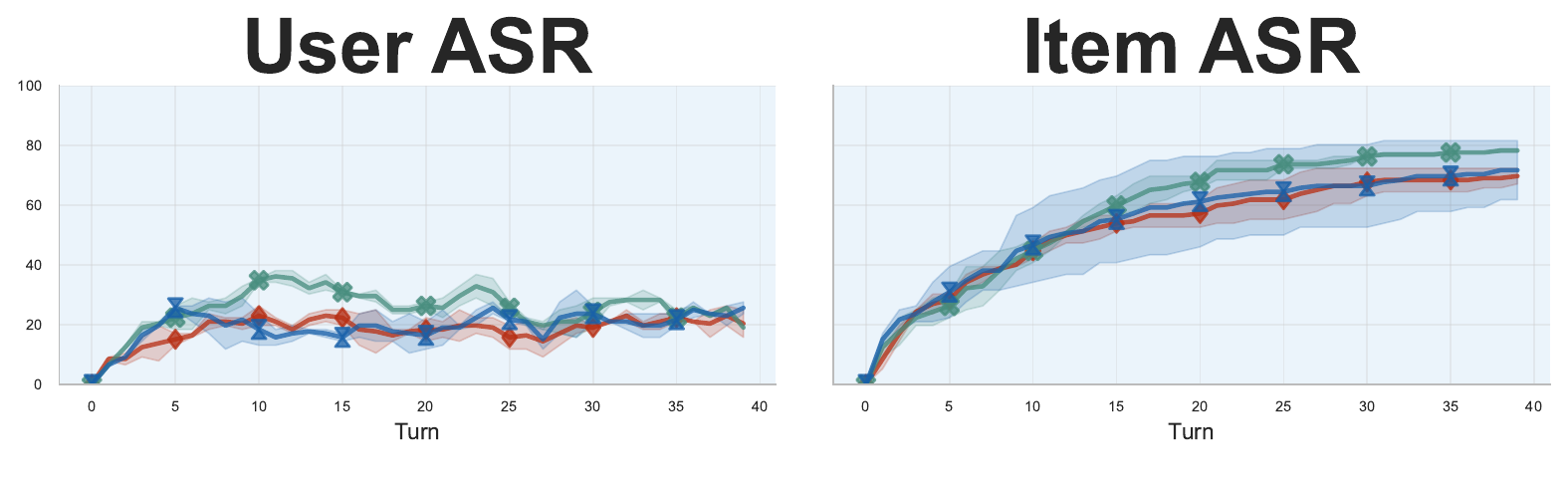}\\[-1.2pt]
  \includegraphics[width=\linewidth]{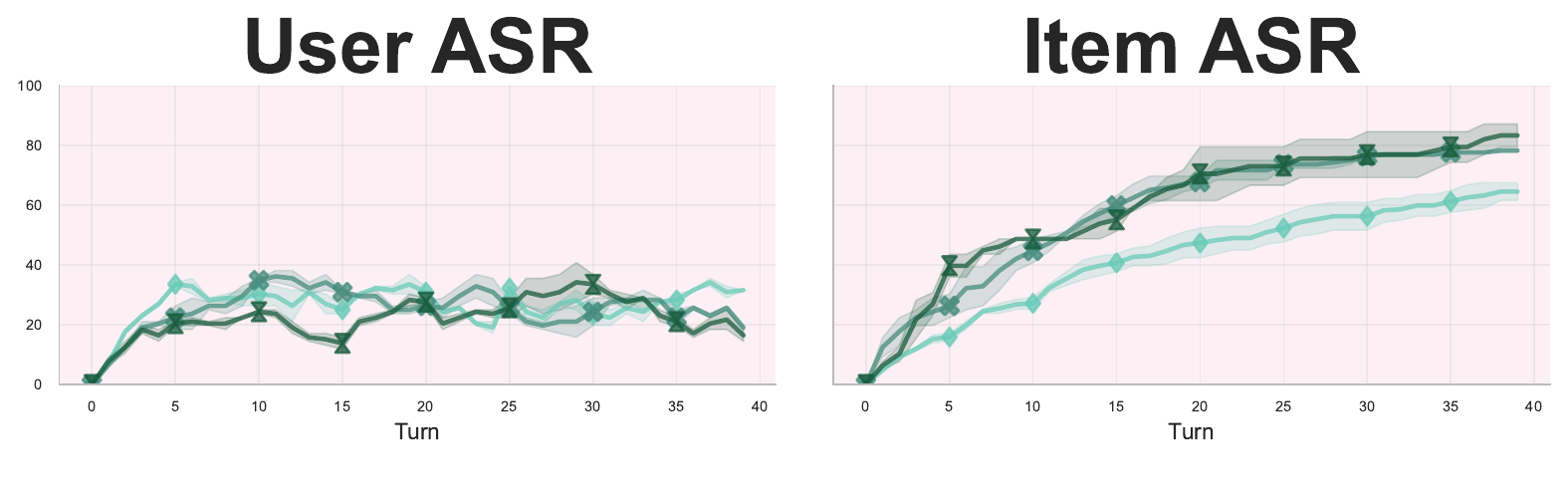}
\end{subfigure}\hfill
\begin{subfigure}[t]{0.24\textwidth}
  \caption{Claude Sonnet 4.5}
  \includegraphics[width=\linewidth]{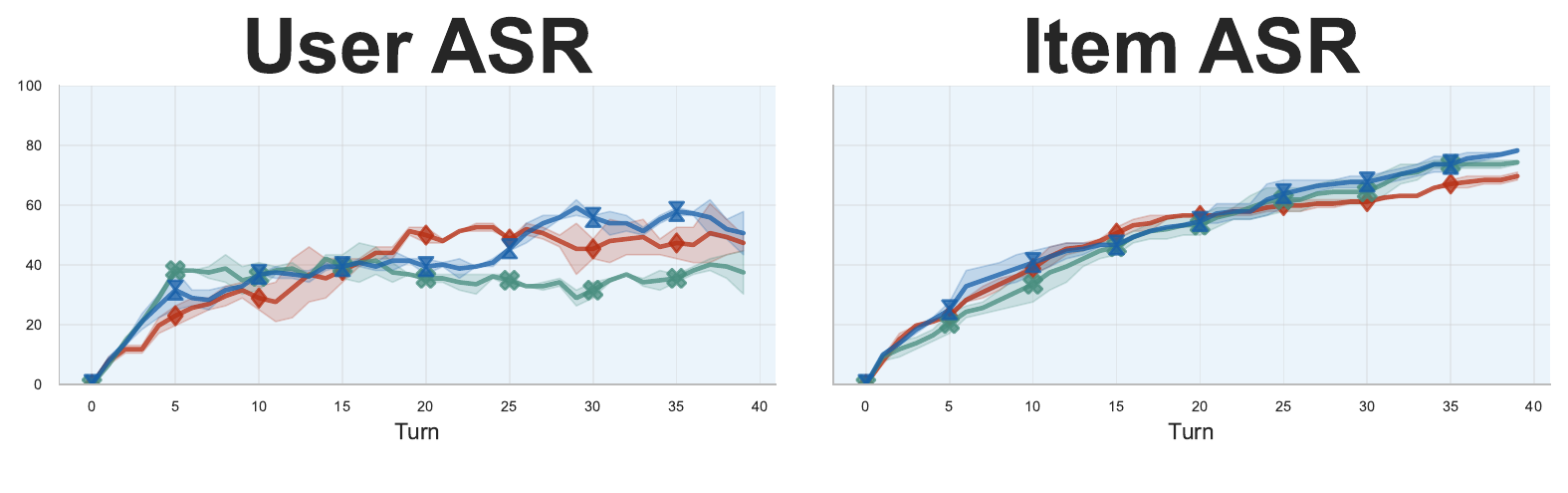}\\[-1.2pt]
  \includegraphics[width=\linewidth]{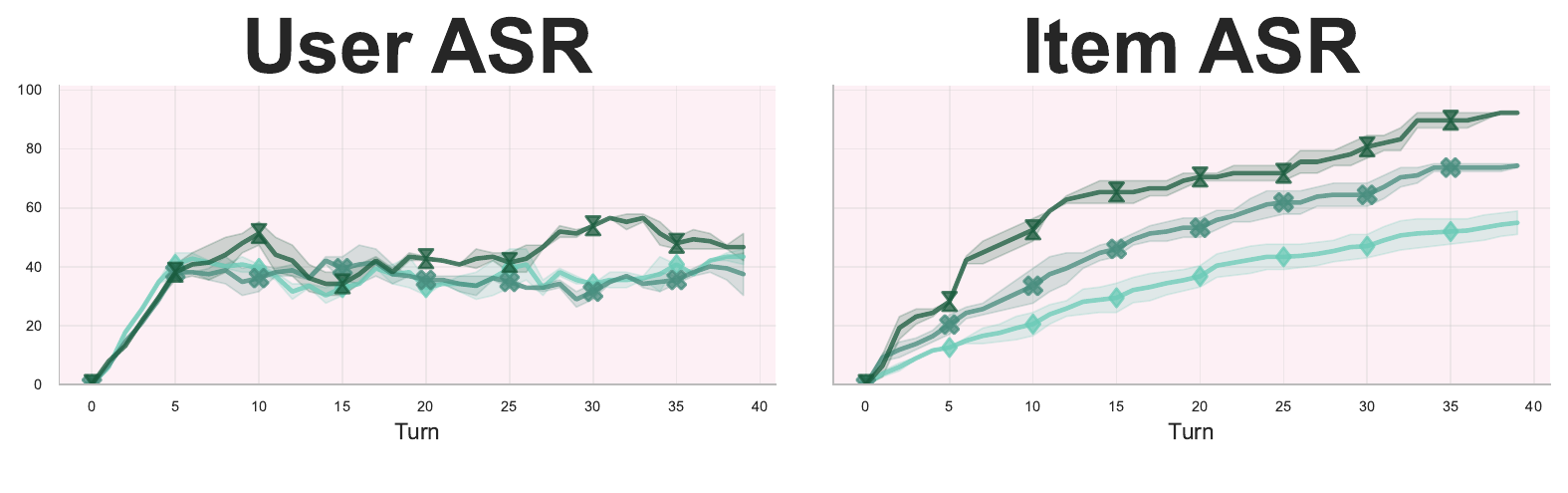}
\end{subfigure}
\hfill
\begin{subfigure}[t]{0.24\textwidth}
  \caption{Mixtral 8$\times$7B}
  \includegraphics[width=\linewidth]{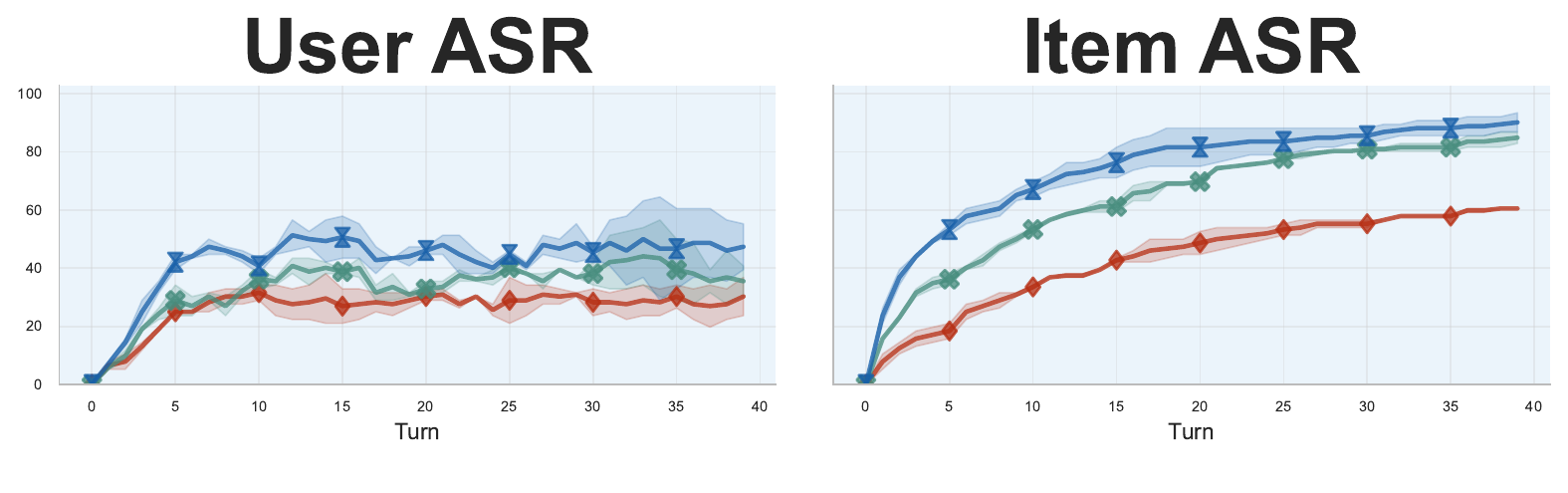}\\[-1.2pt]
  \includegraphics[width=\linewidth]{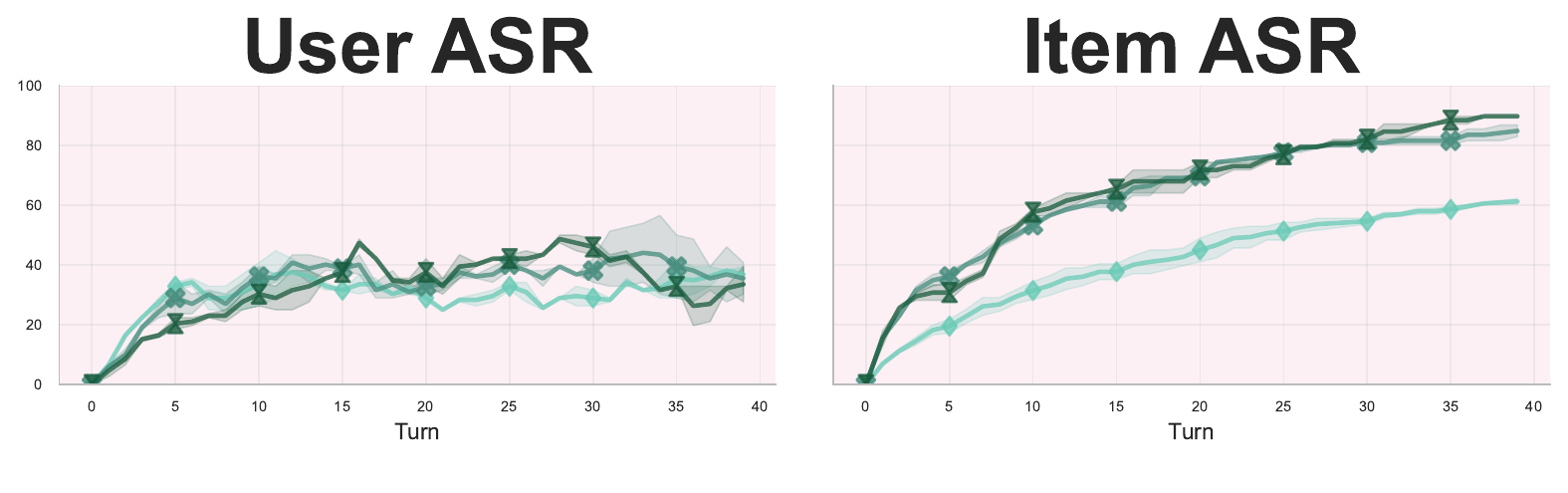}
\end{subfigure}
\hfill
\begin{subfigure}[t]{0.24\textwidth}
  \caption{Llama 4 Maverick 17B}
  \includegraphics[width=\linewidth]{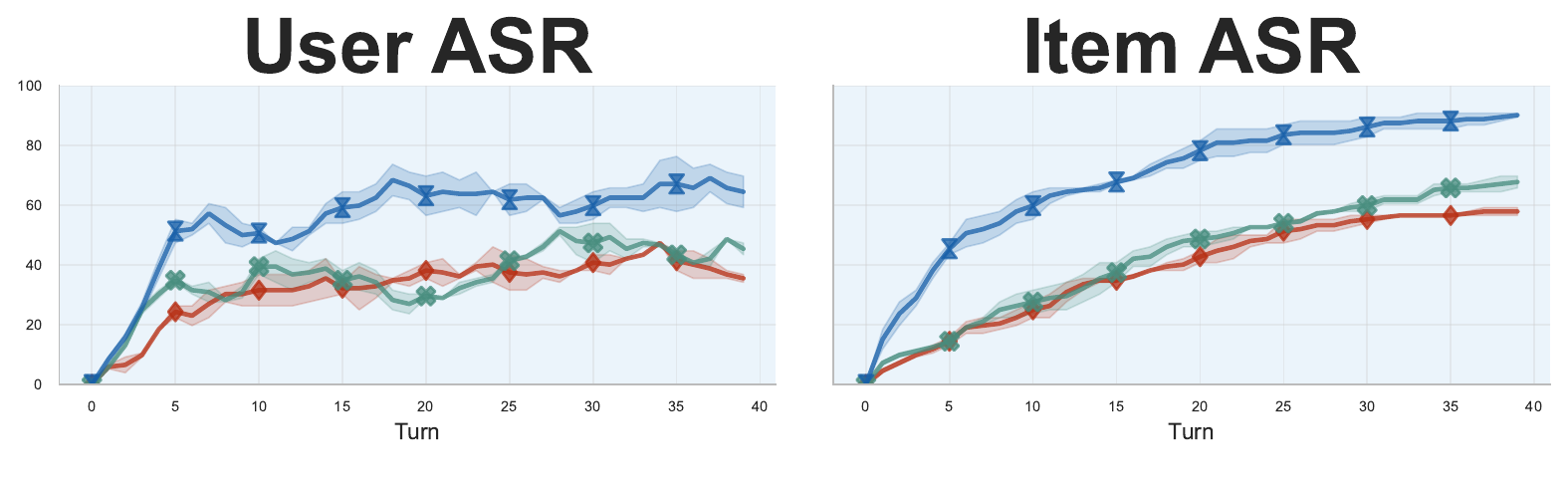}\\[-1.2pt]
  \includegraphics[width=\linewidth]{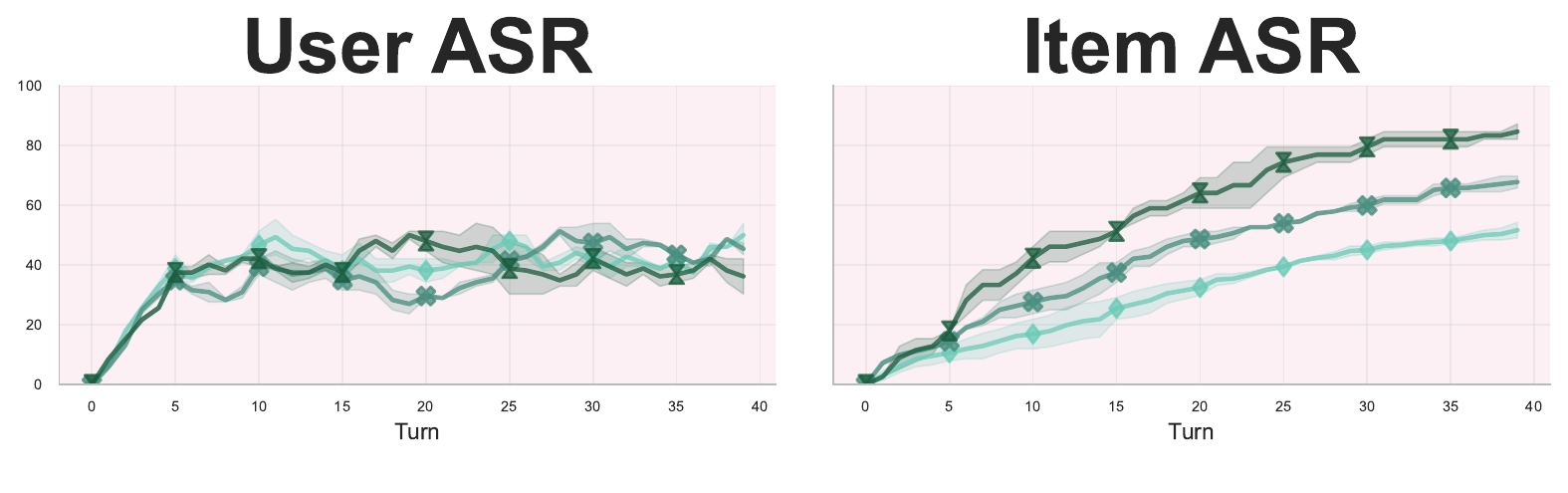}
\end{subfigure}
\vspace{-1.52em}
\caption{Cross-LLM consistency ablation (NetSafe, 25\% attackers). Top: \hlIT{$k$}; bottom: \hlIM{$\rho$}. Claude 4.5 variants exhibit better robustness, likely due to better world knowledge and stronger recovery from benign interaction.}
\label{fig:llm_consistency}
\end{figure*}

\pseudosubsec{Attacker Concentration}
\label{subsubsec:ablation-concentration}
We repeat NetSafe across $(\alpha_U, \alpha_I) \in \{0.10, 0.25, 0.50, 0.75\}^2$. Three takeaways emerge (Fig~\ref{fig:ablation_concentration}):
\textit{(1) Higher attacker concentration shortens the transient regime}: higher $\alpha$ steepens the initial slope and brings the plateau forward.
\textit{(2) Qualitative behaviours remain similar}: users generally exhibit larger variance and turbulence (recovery through benign interactions) than items regardless of $\alpha$.
\textit{(3) Asymmetric partition sensitivity}: higher $\alpha$ yields higher ASR in both role partitions, but user-side contamination is more sensitive to $\alpha$ changes (larger gap between $\alpha$=0.10 and 0.75), whereas item-side contamination varies less across $\alpha$ values.

\pseudosubsec{Attacker Style and Role Split}
\label{subsubsec:ablation-role-split}
We evaluate three RecSys-native attacks that differ in attacker placement and payload style (Fig.~\ref{fig:ablation-rolesplit}) through the lens of NetSafe. Two key observations surface:

\begin{table}[h]
\vspace{-5pt}
\centering\small
\label{tab:attack-design-space}
\setlength{\tabcolsep}{4pt}
\resizebox{0.99\columnwidth}{!}{%
\begin{tabular}{@{}l cc c c@{}}
\toprule
\textbf{Attack} & $\textbf{U}$ & \textbf{I} & \textbf{Payload Style} & \textbf{Injection Surface} \\
\midrule
NetSafe~\cite{yu2024netsafe}            & \checkmark & \checkmark & canary concept via system prompt     & system prompt \\
CheatAgent~\cite{ning2024CheatAgent}   & \checkmark &            & canary concept in user memory        & user profile \\
DrunkAgent~\cite{yang2025drunkagent}   &            & \checkmark & canary concept in item memory        & item profile \\
RecTextAttack~\cite{zhang2024stealthy} &       & \checkmark & lexical perturbation of item text    & item profile \\
\bottomrule
\end{tabular}
}
\vspace{-5pt}
\end{table}

\noindent
\textit{(1) Semantic attacks substantially outperform lexical attacks.}
Under identical attacker placement ($\alpha_U{=}0, \alpha_I{>}0$) and connectivity, DrunkAgent (semantic, Fig.~\ref{subfig:ablation-rolesplit-drunk}) outperforms RecTextAttack (lexical, Fig.~\ref{subfig:ablation-rolesplit-rectextatk}) by a significant margin.
The CF memory update step rewrites profiles as coherent natural-language summaries, silently correcting lexically perturbed text while preserving semantically injected content,
making the latter more durable.
\textit{(2) Role asymmetry in CF propagation.}
DrunkAgent (item-side entry, Fig.~\ref{subfig:ablation-rolesplit-drunk}) contaminates users directly and propagates back to items effectively.
In contrast, CheatAgent (user-side entry, Fig.~\ref{subfig:ablation-rolesplit-cheat}) reaches items directly, but struggles to spread among users via a two-hop path through items. Users seem to be more resilient to higher-order, multi-hop dissemination.

\pseudosubsec{Cross-LLM Consistency}
\label{subsubsec:ablation-llms}
We replicate NetSafe experiments with alternative LLMs in Fig.~\ref{fig:llm_consistency}. Notably, systems built upon newer models are harder to attack, likely due to better alignment, extensive world knowledge, and thus, more chances for reflection-driven self-correction of bias and misinformation during profile updates. More detailed analysis is available in the linked codebase.

\section{Predictive Metrics}
\label{sec:predict}

Simulations of large-scale CF systems are expensive, and it is desirable to rank attack severity across system configurations from graph-structural quantities alone, without running full experiments. 
Our empirical observations suggest that a useful predictor would satisfy the following: (i) it is defined \emph{per role partition} (user-side vs.\ item-side), since the two agent populations exhibit distinct contamination dynamics; (ii) it captures \emph{expected per-turn contact frequency} between attacker and victim agents; and (iii) it accounts for \emph{LLM alignment-driven recovery}, or the tendency of non-attacker victim agents to self-correct through subsequent benign interactions and reflections; (iv) the transient growth slope and the steady-state contamination level should be treated as separate targets.

Such patterns are qualitatively consistent with epidemic spreading on bipartite networks~\cite{pastor2015epidemic}: recovery confers no immunity, so agents re-enter the susceptible state immediately, as in Susceptible-Infected-Susceptible (SIS) dynamics concurrently explored in the general MAS safety community~\cite{zhou2026infaguard,singh2026adversarial,wu2025cowpox,leng2025from,wei2025amemguard}. We adopt a two-population (user/item) homogeneous mean-field approximation since \kUI{} is fixed by configuration and the catalog uniformly resampled, leaving agent degrees near-homogeneous by construction. Per-turn exposure follows the discrete-time contact form of~\cite{gomez2010}. Infection is driven by connectivity-driven exposure at partition-specific susceptibility scales $\beta_U, \beta_I$, offset by a contact-mediated, alignment-driven recovery rate $\gamma$ and a system-level additive adjustment $r$:
\begin{equation}
\resizebox{0.92\columnwidth}{!}{$\displaystyle
\begin{aligned}
  \frac{d\rho_U}{dt} &\approx \beta_U(1-\rho_U)\big[1-(1-\rho_I)^{k}\big] - \rho_U\left[(1-\rho_I)\cdot \mkUI{k}\cdot\gamma + r\right], \\
  \frac{d\rho_I}{dt} &\approx \beta_I(1-\rho_I)\big[1-(1-\rho_U)^{\rho k}\big] - \rho_I\left[(1-\rho_U)\cdot \mkIM{\rho}\!\cdot\!\mkUI{k}\cdot\gamma + r\right]
\end{aligned}
$}
  \label{eq:sis-bipartite}
\end{equation}
where an item agent receives $n_U k/n_I = \rho k$ user contacts per turn against a user agent's $k$.
At steady state ($d\rho_U/dt=0$), we freeze $\rho_I$ at the seeded attacker prevalence $\alpha_I$: a \emph{first-order} (one-hop) approximation in which only seeded attackers transmit, disregarding secondary transmission from contaminated victims. Solving for the steady-state prevalence and reporting the contamination odds gives the \textbf{Recovery-Aware First-Order Connectivity} predictor, each partition's $\beta$ absorbed into its own $\gamma$ and $r$:
\begin{equation}
\resizebox{0.80\columnwidth}{!}{$\displaystyle
  R_{{U}} = \frac{1-(1-\alpha_I)^k}{(1-\alpha_I)\cdot k\cdot\gamma + r}, \quad
  R_{{I}} = \frac{1-(1-\alpha_U)^{\rho k}}{(1-\alpha_U)\cdot \rho \cdot k \cdot\gamma + r}
$}
  \label{eq:reff}
\end{equation}
by symmetric derivation on the item side. 
Correlation results are available in the linked codebase, where $(\gamma, r)$ are empirically fitted per attack and partition to maximise mean Spearman correlation. We observe that predictor-outcome correlations are uneven across attackers' goals, roles, and temporal regimes, suggesting this first-order formulation is an over-simplification and that incorporating higher-order contact dynamics would be necessary for more accurate characterization.

\section{Conclusions and Future Work}
\label{sec:conclusion}

We systematically study how candidate count (\kUI{}) and catalog concentration (\rhoIM{}) shape vulnerability in multi-agent CF systems. Six attacks and four defenses are reproduced in the agentic CF setting and evaluated through the lens of connectivity. The two axes affect user and item partitions asymmetrically, and transient growth differently from steady-state contamination, yielding richer connectivity-vulnerability patterns than in general MAS; each is best tracked as a distinct risk. Practically, (1) more homogeneous user cohorts (\hlIM{$\rho{\uparrow}$}) and higher user fan-out (\hlIT{$k{\uparrow}$}) may amplify individual agent impact and warrant care when scaling or reconfiguring CF systems; (2) while LLM alignment drives recovery from contamination (\S\ref{subsubsec:ablation-llms}), connectivity-aware guardrails remain necessary.
We summarize eight key elements of reproducibility for CF in \S\ref{sec:discussions}, which may help structure safety evaluation metric reporting in future work.

\vspace{3pt}
\noindent\textbf{Limitations.}
(1) \emph{Scale \& Domain:} All experiments are conducted on a 100-user MovieLens cohort; generalisation to larger, sparser, or power-law catalog distributions remains open.
(2) \emph{Potential Confounders:} Adjusting \kUI{} co-varies with prompt and ranking task structures, while resampling for \rhoIM{} shifts item degree distributions. Observed effects are reported under these coupled changes.
(3) \emph{Utility Trade-off:} We focus on safety metrics (ASR, leakage); measuring recommendation quality impact (e.g., HR@$k$, NDCG) under active defenses could complement this view.
(4) More systematic calibration of LLM judges, sensitivity to the judge's underlying model, statistical significance across repeated seeds, and adaptive threat models represent important avenues for future extension.

{
\small
\bibliographystyle{ACM-Reference-Format}
\bibliography{ref}
}

\end{document}